\documentclass[twocolumn, twocolappendix]{aastex631}
\usepackage{natbib}
\usepackage{graphicx}	
\usepackage{color}
\usepackage{amsmath}	
\usepackage{amssymb}	
\usepackage{mathrsfs}
\usepackage{mathrsfs}

\newcommand{\msun}{M_\odot}

\newcommand{\zsun}{Z_\odot}

\newcommand{\msunyr}{M_\odot~{\rm yr}^{-1}}

\newcommand{\JWST}{JWST}
\newcommand{\HST}{HST}
\newcommand{\Spitzer}{Spitzer}
\newcommand{\CFHT}{CFHT}

\newcommand{\target}{CEERS-AGN-z5-1}

\newcommand{\angstrom}{\textup{\angstrom}}

\defcitealias{VB01}{VB01}

\shorttitle{A $z=5$ AGN in CEERS}
\shortauthors{Onoue et al.}
\graphicspath{{./}{figures/}}

\begin{document}

\title{A Candidate of a Least-Massive Black Hole at the First 1.1 Billion Years of the Universe}

\correspondingauthor{Masafusa Onoue, Kohei Inayoshi}
\email{onoue@pku.edu.cn}

\author[0000-0003-2984-6803]{Masafusa Onoue}
\altaffiliation{Kavli Astrophysics Fellow}
\affiliation{Kavli Institute for Astronomy and Astrophysics, Peking University, Beijing 100871, China}
\affiliation{Kavli Institute for the Physics and Mathematics of the Universe (Kavli IPMU, WPI), The University of Tokyo, Chiba 277-8583, Japan}

\author[0000-0001-9840-4959]{Kohei Inayoshi}
\affiliation{Kavli Institute for Astronomy and Astrophysics, Peking University, Beijing 100871, China}

\author[0000-0002-0786-7307]{Xuheng Ding}
\affiliation{Kavli Institute for the Physics and Mathematics of the Universe (Kavli IPMU, WPI), The University of Tokyo, Chiba 277-8583, Japan}

\author[0000-0002-1044-4081]{Wenxiu Li}
\affiliation{Kavli Institute for Astronomy and Astrophysics, Peking University, Beijing 100871, China}

\author[0000-0002-8502-7573]{Zhengrong Li}
\affiliation{Kavli Institute for Astronomy and Astrophysics, Peking University, Beijing 100871, China}

\author[0000-0002-8136-8127]{Juan Molina}
\affiliation{Kavli Institute for Astronomy and Astrophysics, Peking University, Beijing 100871, China}

\author[0000-0002-7779-8677]{Akio K. Inoue}
\affiliation{Waseda Research Institute for Science and Engineering, Faculty of Science and Engineering, Waseda University, 3-4-1, Okubo, Shinjuku, Tokyo 169-8555, Japan}
\affiliation{Department of Physics, School of Advanced Science and Engineering, Faculty of Science and Engineering, Waseda University, 3-4-1, Okubo, Shinjuku, Tokyo
169-8555, Japan}

\author[0000-0003-4176-6486]{Linhua Jiang}
\affiliation{Kavli Institute for Astronomy and Astrophysics, Peking University, Beijing 100871, China}
\affiliation{Department of Astronomy, School of Physics, Peking University, Beijing 100871, China}

\author[0000-0001-6947-5846]{Luis C. Ho}
\affiliation{Kavli Institute for Astronomy and Astrophysics, Peking University, Beijing 100871, China}
\affiliation{Department of Astronomy, School of Physics, Peking University, Beijing 100871, China}

\begin{abstract}

We report a candidate of a low-luminosity active galactic nucleus (AGN) at $z=5$ that was selected from the first near-infrared images of the \JWST\ CEERS project.
This source, named  \target\ at absolute 1450 {\AA} magnitude $M_{1450}=-19.5\pm0.3$, was found via a visual selection of compact sources from a catalog of Lyman break galaxies at $z>4$, taking advantage of the superb spatial resolution of the \JWST/NIRCam images.
The 20 photometric data available from \CFHT, \HST, \Spitzer, and \JWST\ suggest that
the continuum shape of this source is reminiscent of that for an unobscured AGN, and 
there is a clear color excess in the filters where the redshifted H$\beta$+[O~{\sc iii}] and H$\alpha$ are covered.
The estimated line luminosity is $L_\mathrm{H\beta+[OIII]} =10^{43.0}$ erg s$^{-1}$ and $L_\mathrm{H\alpha} =10^{42.9}$ erg s$^{-1}$ with the corresponding rest-frame equivalent width  $\mathrm{EW}_\mathrm{H\beta+[OIII]} =1100$ {\AA} and $\mathrm{EW}_\mathrm{H\alpha} =1600$ {\AA}, respectively.
Our SED fitting analysis favors the scenario that this object is either a strong broad-line emitter or even a super-Eddington accreting black hole (BH),
although a possibility of an extremely young galaxy with moderate dust attenuation is not completely ruled out.
The bolometric luminosity, $L_\mathrm{bol}=2.5\pm0.3 \times10^{44}$ erg s$^{-1}$, is consistent with those of $z<0.35$ broad-line AGNs with $M_\mathrm{BH}\sim 10^6 M_\odot$ accreting at the Eddington limit.
This new AGN population at the first 1.1 billion years of the universe may close the gap between the observed BH mass range at high redshift and that of BH seeds. 
Spectroscopic confirmation is awaited to secure the redshift and its AGN nature.

\end{abstract}


\section{Introduction} \label{sec:intro}

The past two decades were a golden era of high-redshift quasar observations.
Starting from the Sloan Digital Sky Survey \citep{Fan01, Jiang16}, 
$\gtrsim$1000 deg$^2$-class wide-field surveys have identified several hundred quasars in the epoch of cosmic reionization with the current redshift record of $z\sim 7.6$ \citep{Banados18, Yang21, Wang21}.
Follow-up spectroscopic studies have revealed that those quasars are powered by accreting massive BHs with 
masses greater than $M_\mathrm{BH}= 10^9~\msun$, despite their young ages \citep[e.g.,][]{Wu15, Shen19}.
The formation and early growth history of massive BHs is one of the biggest mysteries of  modern astronomy.

Exploring the origin of the massive BHs is still challenging because 
the brightest quasar population is already matured after experiencing rapid accretion episodes and losing the information on their seeding process.
Currently, the least massive BHs that have been spectroscopically confirmed at $z\gtrsim5$ are those with $M_\mathrm{BH}\sim 10^8 M_\odot$ \citep[e.g.,][]{Willott10a, Onoue19}, several orders of magnitudes heavier than the predicted seed mass range \citep[$10^{2-6}M_\odot$; see][for a recent review]{Inayoshi20}.
Therefore, it is crucial to close the gap between the two mass regimes via detecting lower-mass BHs
in epochs as early as possible.

Extensive efforts have been made with the Hubble Space Telescope (\HST) to identify $z\gtrsim6$ AGNs at the very faint end of the quasar/AGN luminosity function (UV magnitude $-22$ or fainter).
No AGNs have been confirmed yet \citep{Morishita20, Fujimoto22, Jiang22}, but with a few candidates at $z=7$--8 \citep{Ishikawa22}.
A naive interpretation of their low success rates is that their point-source selection is not an effective approach to search for low-luminosity AGNs, because host galaxy contamination becomes more severe \citep{Bowler21, Adams22a}.
On the other hand, a fraction of high-redshift quasars seem to have compact ($<1$ kpc) host galaxies, according to recent ALMA observations of luminous $z>6$ quasars \citep{Neeleman21, Walter22}.

Aside from pure AGN surveys,  one possible alternative approach to find low-luminosity AGNs is to use samples of UV-selected galaxies.
At redshift $4<z<6$, the UV luminosity functions of AGNs and Lyman break galaxies (LBGs) intersect with each other at $M_\mathrm{UV} \sim-23$ mag \citep{Ono18, Stevans18, Matsuoka18c, Adams22a, Harikane22a, Finkelstein22b, Li22}.
Spectroscopic follow-up studies of $z\gtrsim4$ LBGs have recently found several candidates of AGNs and galaxy+AGN composite sources at the galaxy-dominated regime, based on detection of broad emission lines and high-ionization emission lines such as He~{\sc ii} and N~{\sc v} \citep{Laporte17, Harikane22a}.

The James Webb Space Telescope (\JWST) has just provided its first infrared  images in July 2022.
The 6.5-meter space observatory is expected to be a game changer in the field of extragalactic observations.
Numerous papers have claimed the first detection of ultra high-redshift galaxies at $z\gtrsim10$, waiting for spectroscopic confirmation \citep[e.g.,][]{Castellano_2022,Donnan_2022, Adams22b, Finkelstein22, Finkelstein22c, Harikane_2022c, Naidu_2022b}.
Likewise, the superb sensitivity of JWST may also enable us to witness rapidly accreting seed BHs far beyond the current samples of high-redshift AGNs. \citep[e.g.,][]{Natarajan_2017, Valiante_2018, Inayoshi22, Goulding22}\footnote{\citet{Ono22} report an  AGN candidate at $z\sim12$ (GL\_z12\_1) based on their morphology analysis of $z\simeq9$--17 galaxy candidates.}.

We here report a candidate low-luminosity AGN at $z=5$, slightly after the end of cosmic reionization.
This candidate was selected with the first imaging dataset of the Cosmic Evolution Early Release Science Survey (CEERS; \citealt{CEERS}), one of the 13 \JWST\   Cycle 1 Early Release Science (ERS) programs.
In Section~\ref{sec:selection}, we present our imaging dataset and parent sample of $z>4$ LBGs.
Section~\ref{sec:result} presents our discovery of a promising AGN candidate and its photometric properties.
Our spectral fitting analysis and constraint on the $z=5$ AGN luminosity function is discussed in Section~\ref{sec:discussion}, followed by our future prospects in Section~\ref{sec:summary}.
All magnitudes quoted in this paper are in the AB system, and are corrected for Galactic extinction \citep{Schlafly11}.
$\Lambda$CDM cosmology is adopted with $H_0=70$ km s$^{-1}$ Mpc$^{-1}$, $\Omega_{\rm M}=0.3$, and $\Omega_\Lambda=0.7$, leading to a scale of 6.28 kpc per arcsec at $z=5$.

\section{Candidate Selection}\label{sec:selection}

\subsection{Sample}\label{sec:sample}
The parent sample of our selection is a catalog of $z=2$--$9$ LBGs compiled by \citet{Bouwens21}.
This sample was obtained from multiple imaging surveys of the \HST. 
The CANDELS Extended Groth Strip (EGS) field is covered by the \JWST\ CEERS program.
A multi-wavelength (0.4--8 \micron) source catalog down to \HST/WFC3 F160W magnitude $26.62$ was made available by \citet{Stefanon17}.

This LBG catalog with multi-wavelength photometry is ideal to characterize the near-infrared (NIR) spectral energy distribution (SED) of $z>4$ LBGs and search for low-luminosity AGNs.
We thus initiated our low-luminosity AGN survey by finding \JWST\ counterparts of those known LBGs.

\subsection{Data Reduction}\label{sec:reduction}
The data we analyze in this paper are the first CEERS images taken by \textit{the Near Infrared Camera} \citep[NIRCam;][]{Rieke05} on 20 June 2022.
The CEERS survey employs seven filters of NIRCam (F115W, F150W, F200W, F277W, F356W, F410M, and F444W) that cover 1--5 \micron.
At the time of writing this paper, the full-band coverage of NIRCam is four pointings (CEERS1, CEERS2, CEERS3, CEERS6) of the NIRCam's $2\arcmin.2 \times 4\arcmin.4$ field-of-views, or 34.5 arcmin$^2$.
The total exposure times for each filter is about 2835 seconds, while F115W has twice longer exposure times.
More details of the survey design and the Epoch~1 NIRCam observations of the CEERS program are presented by the ERS team \citep{Bagley22}.

We downloaded the archival Stage~2 data products from the STScI MAST Portal\footnote{\url{https://archive.stsci.edu/}}.
The Stage~2 images were processed with the JWST pipeline version 1.5.3 with the pipeline mapping file \textsf{jwst\_0942.pmap} except for F410M and F444W.
For those two reddest filters multiple pmap files are used (\textsf{jwst\_0877.pmap} and \textsf{jwst\_0878.pmap} for F410M; \textsf{jwst\_0877.pmap}, \textsf{jwst\_0878.pmap}, and \textsf{jwst\_0881.pmap} for F444W), depending on the fields.
The Stage~2 images were post-processed as follows.
It is known that the version 1.5.3 pipeline has an issue in the imaging background subtraction\footnote{\url{https://github.com/spacetelescope/jwst/issues/6920}}.
We first subtracted the global background with \textsf{photutils}'s \textsf{Background2D} function.
There is a clear horizontal and vertical pattern in the current NIRCam images due to detector readnoise.
The so-called $1/f$ noise was removed with the script provided by the CEERS team\footnote{\url{https://ceers.github.io/releases.html\#sdr1}}.

Those post-processed Stage~2 frames were then stacked by the \JWST\ Stage~3 pipeline.
During the \textsf{Resample} step, we decreased the final pixel sampling size by a factor of two with the drizzle algorithm.
This procedure yields the final pixel sizes of 0.0156 arcsecond per pixel in Short Wavelength filters (F115W, F150W, and F200W) and  0.0315 arcsecond per pixel in Long Wavelength filters (F277W, F356W, F410M, and F444W).
The Stage~3 images were aligned to match the coordinates of unsaturated GAIA sources inside each field-of-view based on its DR3 source catalog \citep{GAIA16}. 
Finally, we referred to the post-flight flux calibration files that contain conversion factors from pixel signals per second to mega Jansky per steradian unit (\textsf{jwst\_nircam\_photom\_0101.fits} for module A, and \textsf{jwst\_nircam\_photom\_0104.fits} for module B).
This process was needed to correct the pre-flight measurements applied in the original images.
The difference of the pre- and post-flight photometric reference files is significant, as is also discussed in \citet{Adams22b}.
With the updated calibration, the F410M and F444W flux densities (in mega Jansky per steradian unit) decrease by 10--20 percent in the two NIRCam detector modules.

We also examined the consistency of photometry between images of \JWST/NIRCam and \HST/WFC3 that cover similar wavelength ranges.
We performed aperture photometry for isolated bright stars inside the current footprints of CEERS.
For \HST, we use the WFC3 images processed by the CEERS team\footnote{CEERS HST data release v1: \url{https://ceers.github.io/releases.html\#hdr1}}.
We use apertures with $4\arcsec.5$ diameter for both NIRCam and WFC3 images for this test.
This aperture size is large enough to accumulate the entire light of a star.
As a result, we found that the NIRCam F115W magnitudes offset from the \HST/WFC3 F125W magnitudes by  $-0.05$ mag.
Likewise, the NIRCam F150W magnitudes have $\sim -0.20$-- $-0.15$ mag offsets with respect to the WFC3 F140W and F160W magnitudes.
Since the two WFC3 filters straddle the transmission curve of NIRCam F150W, this offset cannot be explained by stellar SEDs.
Therefore, we added $+0.05$ mag to our photometry for F115W and $+0.15$ mag for F150W.
The amount of possible offsets for the other filters (F200W, F277W, F356W, F410M, F444M) is uncertain at this stage, because there are no \HST\ filters that match with those NIRCam filters.
The absolute flux calibration that we use has 20\%\ uncertainty in all filters, which is consistent with our offset measurements for F115W and F150W.
We thus added $0.2$ magnitude to the uncertainty of photometry for all NIRCam filters, which dominates the error budget for our LBGs.
Further analyses of the ongoing \JWST\ calibration programs are needed to better perform photometry with NIRCam.

\subsection{Visual Selection of Point Sources}
The NIRCam images of $z>4$ LBGs in the CEERS field were visually inspected by one of the authors (M.~Onoue). 
We focus on compact sources in this study, because we are interested in sources dominated by the central BH radiation.
We note that the UV luminosity function at the faint end (rest-frame UV magnitude $m_\mathrm{UV}\gtrsim 24$ mag, or  $M_\mathrm{UV} \gtrsim -22$ mag in the absolute frame) is dominated by star-forming galaxies \citep{Harikane22a, Bowler21, Adams22a}.
In this sense, our morphology selection is only sensitive to a subsample of AGN populations at high redshift, the host galaxies of which are outshined by the central AGNs or have compact morphology.

The visual inspection of the known $z>4$ LBGs returned approximately 20 compact sources without any noticeable extended components or interacting sources.
We ran  \textsf{SExtractor} \citep{SExtractor} for those compact sources to perform initial photometry.
We require that  the difference between the SExtractor's \textsc{MAG\_AUTO} and aperture magnitudes (\textsc{MAG\_APER}) with $0.\arcsec33$ diameter ($\times5$ image FWHM) be less than $0.3$ mag in the F200W images.
Eight objects satisfy this criterion.

We hereafter present one of the compact sources we found from the selection above, 
\target, or EGSV-9176349491 in the original CANDELS catalog with the photometric redshift $z=4.71$.
This source is the only source among the eight that shows F115W $-$ F356W $> 0.5$ mag (and F115W $-$ F444W $> 1.5$) based on \textsc{MAG\_AUTO}.
This unique color is reminiscent of an unobscured AGN, as we present later in this paper. 
The NIRCam coordinate of \target\ is (R.A., Decl.) $=$ (14:19:17.629, $+$52:49:49.04).
We cross-match this source with the CANDELS catalog  \citep{Stefanon17} to compile their optical-to-NIR photometric information.

\section{Results}\label{sec:result}

\subsection{A Candidate AGN at $z=5$}\label{sec:result_agn}

\begin{figure*}[ht!]
\centering
 \includegraphics[width=\linewidth]{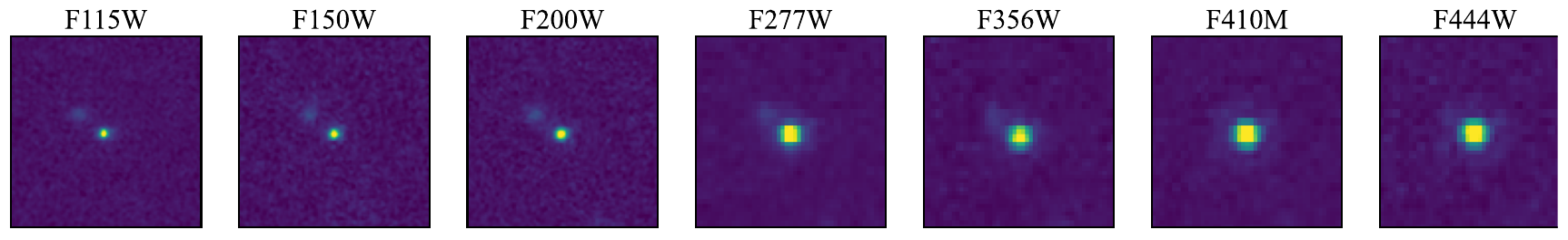}
 \includegraphics[width=\linewidth]{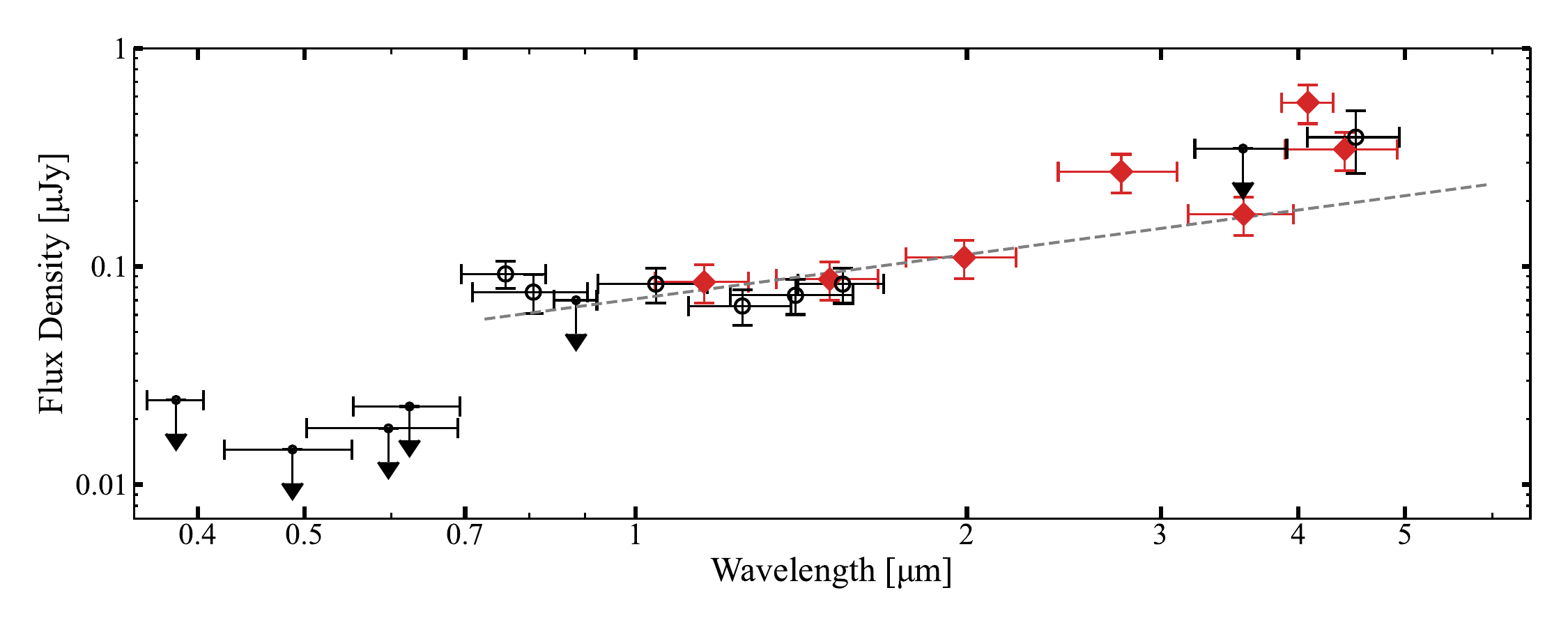}
\caption{
The $z=5$ AGN candidate presented in this paper, \target. 
 (\textit{Top:}) The snapshot images of seven NIRCam filters employed in CEERS. 
 The image size of each panel is $1\arcsec.5 \times 1\arcsec.5$.
 There is a companion source to the upper left from the central source.
 (\textit{Bottom:}) The optical-to-NIR SED of \target.
 The NIRCam flux densities based on model magnitudes are presented in red.
 \target\ has an entry in the CANDELS catalog of \citet{Stefanon17}.
 Here we show optical \CFHT/MegaCam ($u^*$, $g'$, $r'$, $i'$, $z'$), 
  \HST/ACS (F606W, F814W) $+$ WCS3 (F105W, F125W, F140W, F160W), 
  and \Spitzer/IRAC ($3.6~\micron$ and $4.5~\micron$) in black.
   Three sigma upper-limit flux densities are shown for those with signal-to-noise ratios less than $3$.
 Horizontal and vertical error bars correspond to the filter bandwidths and photometric errors, respectively.  
 The grey dashed line shows our best-fit power-law continuum model, where the continuum slope index is $\alpha_\lambda=-1.32\pm0.30$. 
} \label{fig:sed}
\end{figure*}

\begin{deluxetable}{lCC}[htbp!]
\tablecaption{Optical-to-NIR photometry of \target\ and the companion source \label{tab:photometry_HST}}
\tablecolumns{3}
\tablenum{1}
\tablewidth{0pt}
\tablehead{
\colhead{Filter} &
\colhead{\target} &
\colhead{Companion} 
}
\startdata
\JWST/NIRCam F115W & 26.6\pm0.2 & 27.6\pm0.2\\
\JWST/NIRCam F150W & 26.5\pm0.2 & 27.4\pm0.2 \\
\JWST/NIRCam F200W & 26.3\pm0.2 & 27.5\pm0.2 \\
\JWST/NIRCam F277W & 25.3\pm0.2 & 28.9\pm0.2 \\
\JWST/NIRCam F356W & 25.8\pm0.2 & 28.7\pm0.2 \\
\JWST/NIRCam F410M & 24.5\pm0.2 & \cdots \\
\JWST/NIRCam F444W & 25.1\pm0.2 & \cdots \\
\CFHT/MegaCam $u^*$    & >27.9 & \\
\CFHT/MegaCam $g'$     & >28.5 & \\
\CFHT/MegaCam $r'$     & >28.0 & \\
\CFHT/MegaCam $i'$     & 26.5\pm0.2 & \\
\CFHT/MegaCam $z'$     & >26.8 & \\
\HST/ACS F606W & >28.2 & \\
\HST/ACS F814W & 26.7\pm0.2 & \\
\HST/WFC3 F105W & 26.6\pm0.2 & \\
\HST/WFC3 F125W & 26.8\pm0.2 & \\
\HST/WFC3 F140W & 26.7\pm0.2 & \\
\HST/WFC3 F160W & 26.6\pm0.2 & \\
\Spitzer/IRAC $3.6\micron$   & >25.0 & \\
\Spitzer/IRAC $4.5\micron$   & 24.9\pm0.3 &  \\
\enddata
\tablecomments{
The photometric errors for NIRCam filters are from the 20\% uncertainty of the absolute flux calibration.
The \CFHT, \HST/ACS, and \Spitzer\ photometry is from the multi-wavelength catalog of \citet{Stefanon17}.
The \HST/WFC3 magnitudes are updated based on our \textsf{Galight} photometry.
Three sigma upper limits are provided for filters with no detection.
We note that our \textsf{SExtractor} MAG\_AUTO photometry for CFHT $i'$-band and \HST/ACS F814W returns brighter magnitudes than reported here by $\approx0.2$---$0.3$ mag, and we get $2.9$ sigma detection in CFHT $z'$-band. Nevertheless, we use the \citeauthor{Stefanon17}'s values to be consistent with other filters. This difference does not have a major effect on our SED fitting analysis in Section~\ref{sec:result_sedfit}.
}
\end{deluxetable}

\target\ shows a remarkably red NIRCam color among the selected compact LBGs, based on our initial \textsf{SExtractor} photometry.
The unique photometric color was later confirmed by our additional photometry with a two-dimensional image modeling tool \textsf{Galight} \citep{Ding20, Ding21}, which we report in Table~\ref{tab:photometry_HST}.
The photometry with \textsf{Galight} assumes that the central source is a combination of point spread function (PSF) and a two-dimensional S\'{e}rsic profile.
The PSF models applied in each filter are based on a library of point source profiles sampled from the CEERS images.
More details on the \textsf{Galight}  analysis with \JWST\ images are presented in \citet{Ding22}.
In this paper, we use the total magnitudes to characterize the SED of \target\ in the following analysis.
We found that the central pixels in the F115W image are dominated by the PSF,
which indicates that the effective radius is $<0.02$ arcsec, or $<0.13$ kpc.
This upper limit of the source size is well below the $1\sigma$ range of $z\sim5$ LBGs
at the rest-frame UV magnitude of $-20.0$ \citep[0.3--1.0 kpc; ][]{Shibuya15}.
The \CFHT, \HST, and \Spitzer\ magnitudes available for this source are also reported in Table~\ref{tab:photometry_HST},
while for \HST/WFC3 filters we also applied \textsf{Galight} to update the photometry from the catalog.
There is no X-ray counterpart in the Chandra catalog of \citet{Nandra15} within a 5\arcsec search radius.

We note that there is a faint companion source visible in the NIRCam images (Figure~\ref{fig:sed}).
The NIRCam coordinate of this source is (R.A., Decl.) $=$ (14:19:17.628, $+$52:49:48.79), approximately $0.23$ arcsecond to the south of \target.
The 2D flux distribution of this companion is simultaneously fitted with the main source by \textsf{Galight} to deblend the two sources, as we report in Table~\ref{tab:photometry_HST}.
This companion is not identified in the reddest F410M and F444W filters.
The possible blending of the companion source in these red filters is likely a minor issue,
because the companion is $\approx1$ mag fainter than \target\ in the SW filters and the difference becomes larger with $>2$ mag for F277W and F356W.

Figure~\ref{fig:sed} shows the NIRCam cutout images and the optical-to-NIR SED of \target.
Thanks to the wealth of photometry available in the CEERS field, 
the rest-frame UV-to-optical SED of \target\ is clear.
There is a strong Lyman break between \CFHT\ $r'$- and $i'$-band (and \HST/ACS F814W).
The observed continuum redward of Lyman break is red with F115W $-$  F356W $= 0.8 \pm 0.3$.
Moreover, there is a clear color excess at 3 \micron\ (F277W) and 4 \micron\ (F410M, F444W, and IRAC 4.5~\micron), while F356W traces the continuum in between.
This excess matches with the redshifted H$\beta$+[O~{\sc iii}] and H$\alpha$ emission lines, respectively.
The strong excess of F410M suggests that H$\alpha$ emission line is within the F410M coverage.
Those photometric features suggest that \target\ is a $4.9\leq z\leq5.6$ source with emission lines so strong that they affect broad/medium-band photometry, as we discuss in more details later (Sections~\ref{sec:result_BBexcess} and \ref{sec:result_sedfit}).

\subsection{Continuum Properties}\label{sec:result_cont}
The broad-band photometry of \target\ is well reproduced by a continuum model with a single power-law function, except for filters that cover strong emission, especially  H$\beta$+[O~{\sc iii}] and H$\alpha$.
Our best-fit model based on photometry for F115W, F150W, and F200W filters suggests
a power-law slope $\alpha_\lambda$ ($\equiv {\rm d}\ln F_\lambda/ {\rm d}\ln \lambda$)
of $-1.32\pm 0.30$, which is consistent with a typical value for type~1 quasars \citep[e.g.,][hereafter \citetalias{VB01}]{Fan01c, VB01}.
With five more filters from \CFHT\ $z'$-band and \HST/WFC3 (F105W, F125W, F140W, and F160W), the slope gets slightly flatter with a larger error ($\alpha_\lambda=-1.25\pm 0.76$).
In what follows, we adopt the former model with NIRCam only, which is presented in the bottom panel of Figure~\ref{fig:sed} (dashed line).
Note that the different choice of the spectral index causes only $< 7\%$ of the difference in the estimated continuum flux density at rest-frame 3000 and 5100 {\AA}.

Adopting the fitted spectral index, we estimate the absolute magnitude 
at rest-frame 1450~{\AA} as $M_{1450}=-19.5 \pm 0.3$ mag at $z=5$.
The monochromatic luminosity at rest-frame 3000~{\AA} and 5100~{\AA} are 
$\lambda L_{3000}= 4.6\pm0.5 \times 10^{43}$ erg s$^{-1}$ and 
$\lambda L_{5100}= 3.9\pm0.4 \times 10^{43}$ erg s$^{-1}$, respectively.
The rest-frame UV brightness of \target\ is more than three magnitudes fainter than 
those for spectroscopically-confirmed $z\sim5$ quasars from the moderately deep optical survey by Subaru/HSC \citep{Niida20}.
We estimate the bolometric luminosity by applying the bolometric correction for $\lambda L_{3000}$ \citep{Richards06a}\footnote{The bolometric luminosities of the \citet{Richards06a}'s sample are $L_\mathrm{bol}>10^{45}$ erg s$^{-1}$, approximately a half dex higher than that of \target. 
The luminosity dependence of bolometric correction is discussed in, for example, \citet{Netzer19}.
With their prescription, the bolometric correction factors become twice larger for \target.},
which yields $L_\mathrm{bol}=5.15\times \lambda L_{3000} = 2.5\pm0.3 \times 10^{44}$ erg s$^{-1}$.
The expected BH mass is $M_\mathrm{BH} = 2\times 10^6~\msun$ if we assume Eddington-limit accretion.
The inferred bolometric luminosity becomes $L_\mathrm{bol}=4.1 \pm 0.4 \times 10^{44}$ erg s$^{-1}$, when we use the correction factor of $9.26$ 
for $\lambda L_{5100}$  instead.

\begin{figure}
\centering
\includegraphics[width=\linewidth]{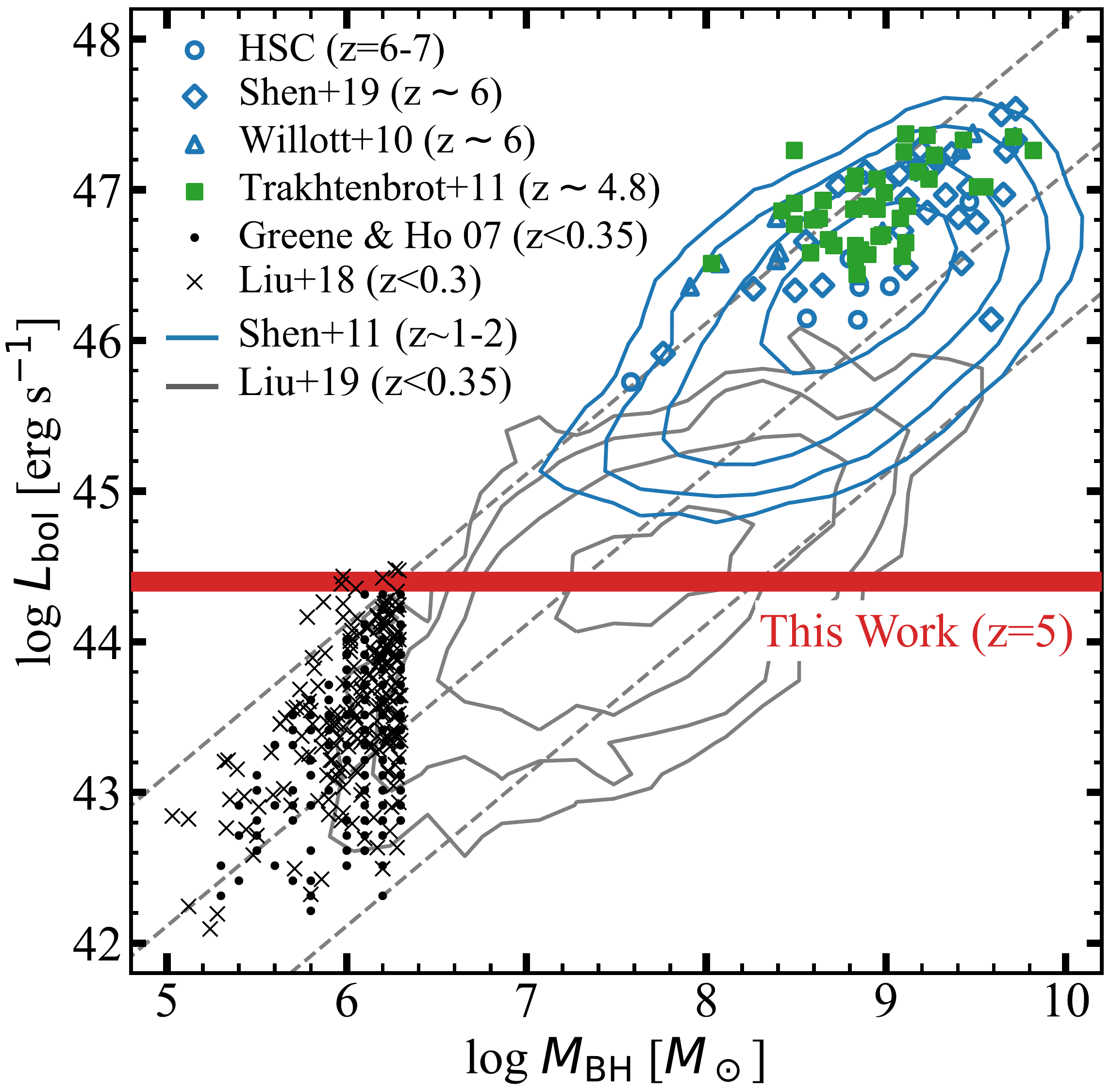}
\caption{
The BH mass - bolometric luminosity plane.
The bolometric luminosity of \target\ ($L_\mathrm{bol}=2.5 \times 10^{44}\  \mathrm{erg~s^{-1}}$; red line) is estimated from its 3000 {\AA} monochromatic luminosity.
The virial BH masses of broad-line AGNs at various redshift ranges are also shown for comparison.
Blue symbols are $z>6$ quasars from Subaru/HSC \citep[circle;][]{Onoue19, Matsuoka19a, Kato20},
SDSS \citep[diamond;][]{Shen19}, and CFHQS \citep[triangle;][]{Willott10a}.
Green squares show $z\sim4.8$ quasars from \citet{Trakhtenbrot11}.
The blue contour is the normalized distribution of SDSS DR7 quasars with a logarithm step of 0.5 dex. \citep{Shen11}.
The BH masses from the literature above are estimated based on Mg~{\sc ii} $\lambda2798$.
The grey contour shows the normalized distribution of low-redshift ($z\lesssim0.35$) broad-line AGNs \citep{Liu19} with the same step as for the \citeauthor{Shen11}'s $z\sim1$--$2$ distribution.
Black symbols are the individual low-redshift AGNs with estimated BH masses
$M_\mathrm{BH}\lesssim10^{6.3}M_\odot$ from \citet[][dot]{GH07c}, and \citet[][cross]{Liu18}.
Those low-redshift samples use Balmer lines to estimate BH masses.
Clearly, \target\ has the typical luminosity of the $z<0.35$ AGNs with $M_\mathrm{BH}\gtrsim10^{6}M_\odot$.
The three diagonal lines indicate 100\%, 10\%, and 1\% Eddington luminosity from top left to bottom right.
} \label{fig:MBH}
\end{figure}

Figure~\ref{fig:MBH} shows the BH mass - bolometric luminosity plane for AGNs at different redshift ranges.
The inferred bolometric luminosity of \target\ is more than two dex smaller than those of the typical $z=1$--2 SDSS DR7 quasars with $L_\mathrm{bol}\simeq 10^{46.5}$ erg s$^{-1}$ \citep{Shen11} and those of known $z\gtrsim5$ quasars with virial BH masses available from Mg~{\sc ii} \citep{Willott10a,Trakhtenbrot11, Onoue19, Matsuoka19a,Kato20}.
The luminosity of \target\ is rather comparable to those for typical low-redshift ($z<0.35$) broad-line AGNs \citep{Liu19}.
In Figure~\ref{fig:MBH}, we show that some $M_\mathrm{BH} \sim10^6~\msun$ BHs from \citet{GH07c} and \citet{Liu18} have comparable luminosities to \target.
An even less massive BH can power this source if rapid super-Eddington accretion is achieved.
The necessity of such intermittent super-Eddington phases has been recently argued in theoretical predictions of early BH assembly \citep{Inayoshi_2022a,Hu_2022b,Shi_2022}.
On the other hand, the same luminosity can also be achieved by a sub-Eddington BH with $M_\mathrm{BH} \simeq 10^{7-8}~\msun$,
as long as the nuclear accretion disk settles down to a radiatively efficient state ($L_{\rm bol}/L_{\rm Edd}\gtrsim 0.01$; see \citealt{Yuan_Narayan_2014}).
Spectroscopic follow-up observations of broad Balmer lines or other virial BH mass tracers such as Mg~{\sc ii} $\lambda2798$ are necessary to robustly estimate the BH mass of \target.

\subsection{Broad/Medium-Band Excess}\label{sec:result_BBexcess}

We now quantify the broad/medium-band excess due to ${\rm H}\beta$+[O~{\sc iii}] and ${\rm H}\alpha$.
Here we consider the NIRCam photometry of F277W, F410M, and F444W, and use our continuum model with a single power-law index $\alpha_\lambda=-1.32$ (Sec.~\ref{sec:result_cont}). 
The observed F277W magnitude is $0.7$ mag brighter than that expected from the continuum flux at F277W, which is significantly larger than the photometric error ($0.2$ mag).
This excess is consistent with a luminosity for the H$\beta$+[O~{\sc iii}]  lines of  $L_\mathrm{H\beta+[OIII]} =10^{43.0}$ erg s$^{-1}$ and rest-frame equivalent width  ${\rm EW}_{\rm H\beta+[OIII]}=1100~{\rm \AA}$.
The H$\alpha$ excess in F410M and F444W, 1.2 and 0.6 mag, respectively, is explained by a strong H$\alpha$ emission with line luminosity  $L_{\rm H\alpha}=10^{42.9}$ erg s$^{-1}$ and rest-frame equivalent width ${\rm EW}_{\rm H\alpha}=1600~{\rm \AA}$.
In addition, assuming the relation of $L_{\rm H\alpha}=3.1~L_{\rm H\beta}$ expected for Case B' recombination \citep[e.g.,][]{GH05},
we infer the [O~{\sc iii}] luminosity as $L_{\rm [OIII]}\simeq 10^{42.9}$ erg s$^{-1}$.
Note that these measurements slightly increase by $5$ and 1 \% for H$\beta$+[O~{\sc iii}] and H$\alpha$, respectively, when the second continuum model for $\alpha_\lambda=-1.27$ is used.

The two equivalent widths for H$\beta$+[O~{\sc iii}] and ${\rm H}\alpha$ are 
extremely large as an AGN.
The composite spectrum of low-redshift quasars of \citetalias{VB01} shows ${\rm EW}_{\rm H\beta+[OIII]}\simeq 63~{\rm \AA}$ and ${\rm EW}_{\rm H\alpha}\simeq 195~{\rm \AA}$.
The brightest type~2 quasars in the local universe show equivalently strong [O~{\sc iii}] line emission (both in terms of luminosity and equivalent width)  \citep[e.g.,][]{Zakamska_2003, KH18},
while their bolometric luminosity estimated from the extinction-corrected [O~{\sc iii}] luminosity $L_\mathrm{bol}\sim 10^{47}~{\rm erg~s}^{-1}$ is $\gtrsim 100$ times higher than that of \target~\citep{Heckman_2004}.
We also note that the estimated H$\alpha$ luminosity of \target\ is five times higher than expected from the empirical relation  between the  H$\alpha$ luminosity and the 5100 {\AA} luminosity for low-redshift broad-line AGNs \citep[][their equation~1]{GH05}.

Alternatively, strong Balmer emission lines can also be produced from star-forming galaxies.
Such high values of $\mathrm{EW}_\mathrm{H\alpha}\simeq 1000~\mathrm{\AA}$ have been reported in 
$z>6$ star-forming galaxies based on \Spitzer\ and \JWST\ \citep[e.g.,][]{Endsley21, Chen22, Stefanon_2022}.
However, those galaxies show a significantly steeper continuum slopes ($\alpha_\lambda\simeq -2.0$),
compared to that of \target.
We will discuss the possible contribution of star-forming galaxies to the SED of \target\ in
Section~\ref{sec:result_sedfit}.

\section{Discussion}\label{sec:discussion}

\subsection{SED Fitting}\label{sec:result_sedfit}
The available photometry of \target\ continuously covers the SED from Ly$\alpha$ to H$\alpha$.
We here present our SED fitting analysis with 
templates of metal-poor galaxies, low-redshift quasars, and super-Eddington accreting BHs.
The redshift range is limited to $4.0\leq z \leq 6.0$ (with steps of 0.01), which is wide enough to cover the redshifted H$\alpha$ emission with F410M.
Intergalactic medium absorption is taken into account with the prescription of \citet{Madau95}.

\begin{figure*}
\centering
 \includegraphics[width=0.8\linewidth]{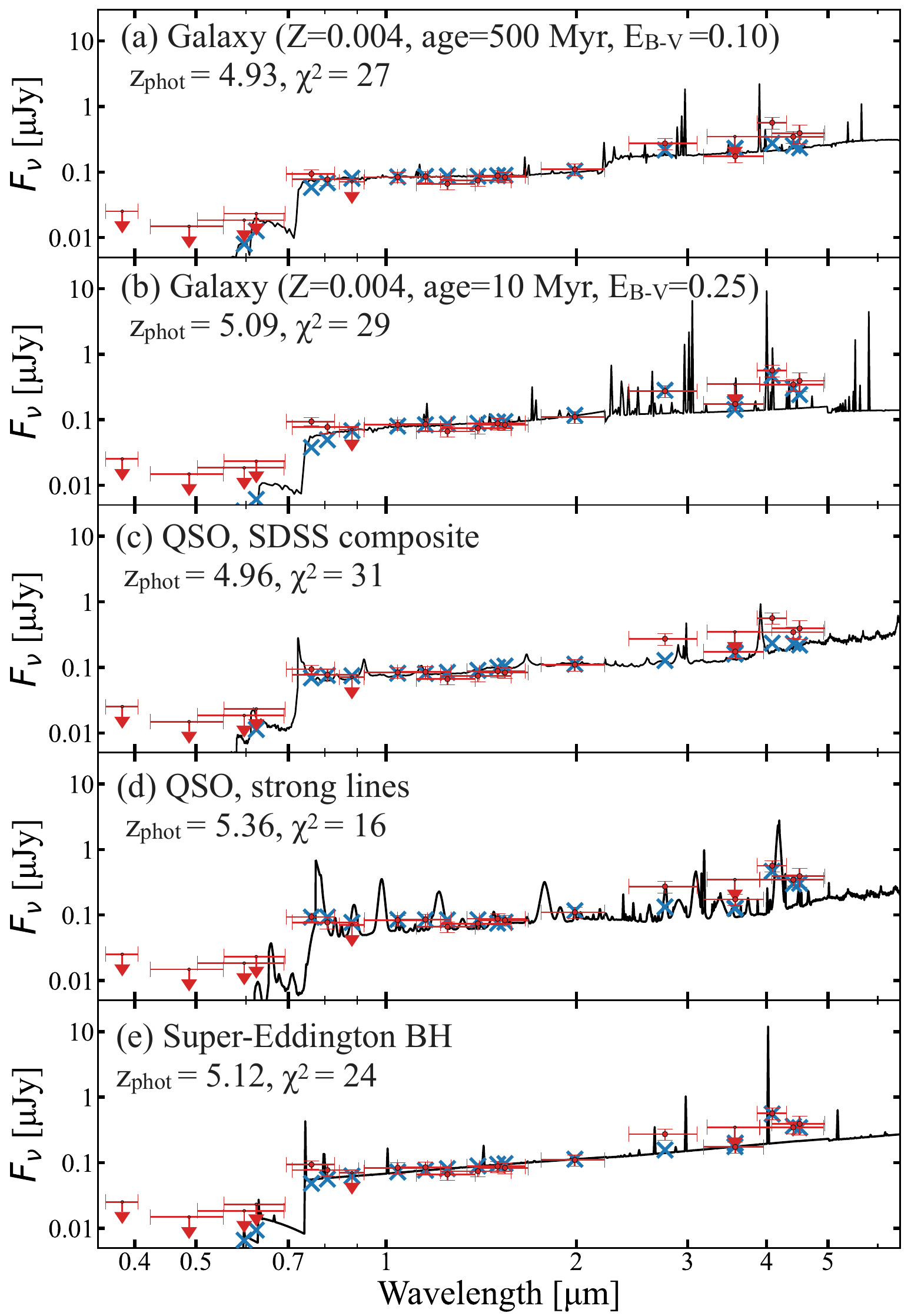}
\caption{
The results of our SED fitting. 
In each panel, we show the best-fit model of galaxies or AGNs with a black line.
The photometric redshift and the chisquare of the best-fit model are shown at the upper left of each panel.
The observed photometry is shown in red dot and the filter-convolved magnitudes of the best-fit models in blue cross.
(a) 
The best-fit galaxy template from \citet{Inoue11}'s galaxy models.
This model is a metal-poor ($Z=0.004$) galaxy at $z=4.93$ with stellar age of 500 Myr, including dust attenuation of $E(B-V)=0.1$ mag with the Calzetti law.
(b) The best-fit galaxy model when the stellar age is fixed to 10 Myr.
This model is also a $Z=0.004$ galaxy with photometric redshift $z=5.09$.
Heavier dust attenuation ($E(B-V)=0.25$ mag) by the Calzetti law is required to trace the continuum.
(c) The low-redshift composite quasar spectrum of \citet{VB01} scaled to match the photometry of \target.
(d) The same low-redshift composite spectrum with emission lines added to match H$\alpha$.
See the text for the details of the procedure.
(e) The SED model of an accreting super-Eddington BH \citep{Inayoshi_2022a}.
The scale of the best-fit model corresponds to the central BH mass of $M_\mathrm{BH}=10^{6.3}M_\odot$.
} \label{fig:SEDfit}
\end{figure*}

\subsubsection{Metal-Poor Galaxies} \label{sec:result_sedfit_galaxy}
First, we attempt to reproduce the observed photometric SED  with galaxy templates.
We here adopt the galaxy SED models from \citet{Inoue11}, where nebular emission lines are included 
for various metallicities of $Z=0.02~(\zsun)$, $0.008$, $0.004$, $0.0004$, 
$10^{-5}$, $10^{-7}$, and $0.0$ (no metals).
The stellar mass spectrum is calculated by assuming a Salpeter initial mass function (IMF) with $1$--$100~\msun$.
We use models of constant star formation history with ages of 10, 100 and 500 Myr, while we test galaxy models with extremely young ages ($<10$ Myr) later in Section~\ref{sec:caveats}.
We take into account dust attenuation by extinction laws 
of the Small Magellanic Cloud (SMC; \citealt{Prevot_1984}) and of starburst galaxies \citep{Calzetti00}.

Figure~\ref{fig:SEDfit}(a) shows our best-fit model at $z=4.93$ (chisquare $\chi^2=27$)
with a metallicity $Z=0.004$, stellar age  
$500$ Myr, star formation rate (SFR) $3.6~\msunyr$, and color excess $E(B-V)=0.10$ mag with the Calzetti extinction law.
This model can well explain the observed continuum owing to modest dust attenuation.
However, the model hardly reproduces the observed flux densities of the filters affected by H$\beta$+[O~{\sc iii}] (F277W) and H$\alpha$ emission lines (F410M, F444W, and IRAC 4.5~\micron),
while they are slightly raised by active starbursts.

We also show in Figure~\ref{fig:SEDfit}(b) the best-fit case when the age is fixed to 10 Myr ($Z=0.004$ and SFR $= 27 M_\odot \mathrm{yr}^{-1}$).
This model can better explain the photometric excess by  H$\beta$+[O~{\sc iii}] and H$\alpha$ emission lines than the previous galaxy model with $500$ Myr age.
However, this model requires heavier extinction of $E(B-V)=0.25$ mag with the Calzetti law, which leads to a poorer fit around Ly$\alpha$.
Nonetheless, the goodness-of-fit becomes just slightly worse ($\Delta \chi^2=2$).

For those galaxy models, it is worth noting the dependence of the fitting goodness on the choice of dust extinction/attenuation laws.
Since the SMC law has a steeper slope than the Calzetti law in the rest-frame UV wavelengths, it gets more difficult to match the galaxy models to the observed SED, which has a flat slope around Ly$\alpha$.
As a result, the best-fit galaxy model with the SMC law is the same as what we show in Figure~\ref{fig:SEDfit}(a) (i.e., 500 Myr age with a milder extinction of $E(B-V)=0.05$ mag), but returns a poorer fit ($\Delta \chi^2=7$).
The disagreement between the observations and the models becomes more serious when only the galaxy models with 10 Myr age are considered, because the SEDs of young galaxies need to be more strongly obscured.
One has to consider $E(B-V)\simeq0.30$ mag to fit the young galaxy model to the rest-optical spectrum with the SMC law, which yields a
substantially worse fit than the model presented in Figure~\ref{fig:SEDfit}(b) ($\Delta \chi^2\simeq50$) owing to the mismatch at the rest-frame UV wavelengths.

\subsubsection{AGNs with Strong Emission Lines} \label{sec:sedfit_AGN_add}

Next, we test the composite spectrum of low-redshift quasars compiled by \citetalias{VB01}.
This composite spectrum from the SDSS covers the rest-wavelength 800--8555~{\AA} and includes
80 resolved emission-line features in the spectrum.
The continuum index measured at rest-frame 1300--5000~{\AA} is $\alpha_\lambda=-1.56$.
Figure~\ref{fig:SEDfit}(c) shows the scaled spectrum at $z=4.96$ matched to \target\ ($\chi^2=31$).
This model agrees with the observed continuum of \target\ well; however, the broad/medium-band excess due to H$\beta$+[O~{\sc iii}] and H$\alpha$ is not sufficiently reproduced (Section~\ref{sec:result_BBexcess}).

The composite spectrum of \citetalias{VB01} is constructed based on luminous quasars,
unlike \target.
It is observationally known that the strength of quasar emission lines increases as the continuum luminosity decreases \citep{Baldwin77}.
To mimic the so-called Baldwin effect, we increase the strength of {\it all} the emission lines listed in Table~2 of \citetalias{VB01},
so that the relative flux ratio of each line is maintained\footnote{We do not increase the strength of the unresolved iron pseudo-continuum in the original spectrum.}.
We add those additional emission line fluxes to the original  spectrum assuming Gaussian profiles with the same line widths as those measured in \citetalias{VB01}.
This modified version of the \citetalias{VB01}'s quasar SED model is presented in  Figure~\ref{fig:SEDfit}(d), in which case a  better goodness-of-fit is achieved at $z=5.36$ ($\chi^2=16$) than the original composite spectrum.
The excess of Ly$\alpha$ and H$\alpha$ is better reproduced than the original model in this case, however the observed flux density at F277W is still twice higher than predicted from this third model.

\subsubsection{Super-Eddington Accreting BHs}  \label{sec:sedfit_AGN_seed}

As an alternative scenario, we consider more theory-based SED model for 
a super-Eddington accreting BH with $M_{\rm BH}\sim 10^6~\msun$ \citep{Inayoshi22}.
The SED model is constructed by post-process line transfer calculations with CLOUDY (C17; \citealt{Ferland_2017}), which is applied to the results of radiation hydrodynamical simulations of a seed BH rapidly growing in the proto-galactic nucleus \citep{Inayoshi_2022a}.
In this model, several prominent lines are powered by the central BH fed via a dense accretion disk at super-Eddington rates.
Among them, strong H$\alpha$ line emission with a rest-frame  ${\rm EW}_{\rm H\alpha}\simeq 1300~{\rm \AA}$ 
is so prominent that the line flux affects the broad-band colors significantly, just like we observe in \target.

Figure~\ref{fig:SEDfit}(e) shows the best-fit model of a super-Eddington accreting BH ($\chi^2=24$) scaled to $z=5.12$.
This model explains the continuum flux densities and the broad/medium-band excess in F410M, F444W, and IRAC 4.5~$\micron$, owing to the strong H$\alpha$ emission powered by the fast-growing BH.
The overall flux normalization of the best-fit case is only 1.4 times higher compared to the original SED model,
suggesting that the predicted line information (e.g., low-ionization emission lines of 
O~{\sc i}~$\lambda\lambda 1304$ and C~{\sc ii}]~$\lambda 2326$) is still valid and can be tested by follow-up spectroscopic observations of \target.
This model yields a slightly larger $\chi^2$ value compared to the quasar model with enhanced emission lines (Fig.~\ref{fig:SEDfit}(c)),
but we note that the SED model for a super-Eddington BH does not include the broad-line components
that could be produced in the inner region but are unresolved in their simulations.

\subsubsection{Caveats}\label{sec:caveats}
Based on our SED fitting analysis, we propose that there are two plausible solutions to explain the observed SED for \target: 
an unobscured AGN with strong broad-line emission (Fig.~\ref{fig:SEDfit}(d)) and
a low-mass super-Eddington accreting BH (Fig.~\ref{fig:SEDfit}(e)).
Those two models well reproduce the continuum shape and the H$\alpha$ excess, both of which originate from the nuclear region of the AGN.
On the other hand, let us recall that
neither of the two models can explain the broad-band excess of F277W, which is likely dominated by [O~{\sc iii}] emission.
Since this forbidden line is not efficiently produced from the dense gas at the nuclear region,
we hypothesize that the F277W excess is at least partly attributed to
the ionized gas at the host-galaxy scale.
For example, a metal-poor galaxy with age $\simeq10$ Myr can produce strong [O~{\sc iii}] emission required to explain the observed magnitude of F277W, when ${\rm SFR}\simeq 30~\msunyr$ (see Figure~\ref{fig:SEDfit}(b)).
We note that this order estimate does not consider the contribution of the host to other filters.

Alternatively, a top-heavy stellar IMF expected in metal-poor environments can also be another solution to produce strong [O~{\sc iii}] emission lines from galaxies, which is not taken into account in the galaxy models we use in this paper.
It is also possible that the nebular gas is oxygen-rich in low metallicity relative to the solar value, which enhances the strength of [O~{\sc iii}] emission without changing the value of metallicity.
The detailed modeling of the additional galaxy SED is left for future investigation.

One may also consider a young galaxy with age  $\lesssim 10$ Myr without AGN contribution, as is discussed in Section~\ref{sec:result_sedfit_galaxy} and also in the literature \citep[e.g.,][]{Inoue11,Wilkins_2020,Tang_2021}.
Indeed, observations with \HST, \Spitzer\ and \JWST\  have suggested that the photometric SEDs of some $z\gtrsim 6$ 
galaxies are consistent with those of young stellar populations down to $\sim3$ Myr ages
\citep[e.g.,][]{Tamura19, Chen22, Endsley21, Endsley22}.
We additionally conduct the SED fitting analysis extending the stellar age down to 1~Myr with delayed star-formation history and
find that a model of a metal-poor galaxy with a metallicity $Z=0.004$, stellar age $1.4$ Myr,
and ${\rm SFR}\simeq 400~\msunyr$ yields the smallest value of $\chi^2=21$ among the galaxy models.
This model specifically needs dust extinction following the Calzetti's law
with color excess $E(B-V)=0.22$ mag.
This level of dust attenuation produces the infrared luminosity of $\sim10^{11.3} L_\odot$, which is converted to the dust mass of $\sim 10^7~\msun$
by assuming dust temperature of $T_{\rm d}=40$ K (see Equation~5 of \citealt{Inoue20}).
However, this amount of dust is hardly produced by the ongoing active star formation within $\approx 1$ Myr
unless a pre-existing stellar population leaves dust grains, and their  older ($\gg 1$ Myr) stellar components are hidden
in the observed SED \citep[e.g.,][]{Tamura19}.

\subsection{$z=5$ AGN Luminosity Function}\label{sec:qlf}

The discovery of the promising candidate of a high-redshift AGN was unexpected from the first CEERS dataset.
To quantify the serendipity, we discuss the number density of $z\sim5$ AGNs based on \target.
We caution that we can only provide a lower bound of the AGN luminosity function with this work
especially because we do not consider AGNs embedded in their host galaxies with extended morphology. 
Such a population would be common for faint AGNs at UV magnitudes dominated by star-forming galaxies.

\begin{figure}
\centering
 \includegraphics[width=\linewidth]{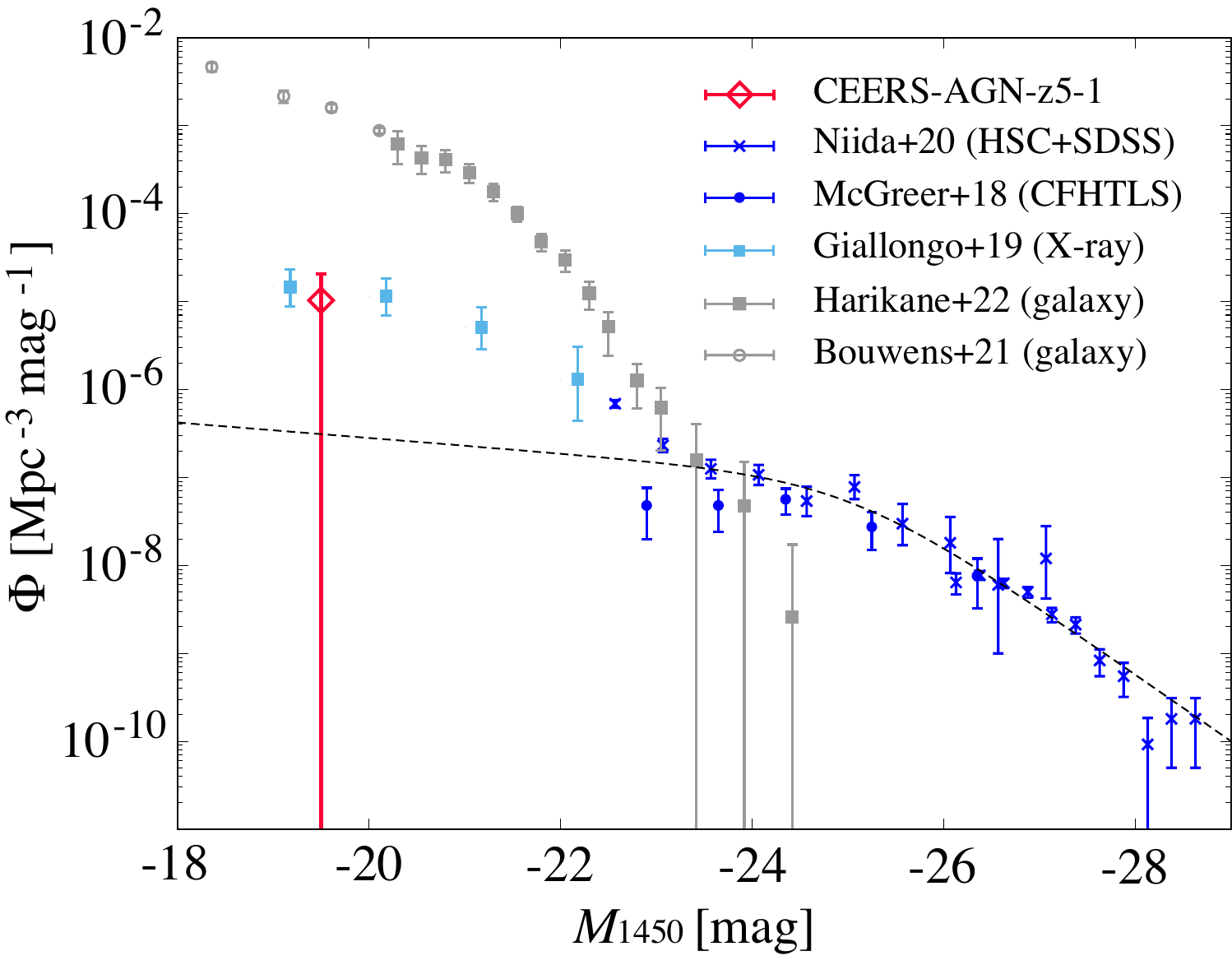}
\caption{
The $z\sim5$ UV luminosity function of AGN.
Our constraint from \target\ is shown in red.
The quasar luminosity function data obtained from different surveys are shown:
the rest-UV selected quasars combining Subaru HSC and SDSS \citep[][cross]{Niida20} and CFHTLS \citep[][dot]{McGreer_2018} in blue.
The abundance of AGN at $M_\mathrm{1450}=-19.5$ mag is significantly higher than the extrapolation of 
the rest-UV selected QLF (dashed line), while within the Poisson error from one object.
Our data point is consistent with the AGN luminosity function from the X-ray detected AGN candidates in the CANDELS field \citep{Giallongo_2019}, which are shown in cyan.
Shown in grey are the UV luminosity function of galaxies from  \citet[][circle]{Bouwens21} and \citet[][square]{Harikane22a}.
} \label{fig:qlf}
\end{figure}

Figure~\ref{fig:qlf} shows our estimate of the $z\sim5$ AGN luminosity function.
We calculate the binned number density as $\Phi = 1.03 \times 10^{-5}~{\rm Mpc}^{-3}~{\rm mag}^{-1}$ at $M_{1450}=-19.5$ mag (red), where we adopt the survey area of $34.5$ arcmin$^2$ and set $z=5.15$ and its interval of $\Delta z=\pm 0.5$ to calculate the cosmic volume.
For comparison, we overlay the binned quasar luminosity functions  derived from 
Subaru/HSC+SDSS \citep{Niida20}, and CFHT Legacy Survey \citep{McGreer_2018}.
We also show the $z=4.5$ X-ray detected AGN luminosity function in the CANDELS fields (including EGS), which go down to $M_{1450}=-18.5$ mag \citep{Giallongo_2019}\footnote{\target\ is not a part of their AGN candidates.}.
In Figure~\ref{fig:qlf}, we take the face values of their $z=4.5$ luminosity function without correcting for redshift evolution.
The abundance of the faintest AGNs at $M_{1450}=-19.5$ mag is substantially higher than that 
expected from extrapolation of the HSC+SDSS QLFs (short dashed; \citealt{Niida20}), while the extrapolation of their double power-law LF is still consistent with our data point within the Poisson error of one object. 
Intriguingly, our binned luminosity function is rather consistent with those of Chandra-detected (0.5--2 keV) X-ray sources from  \citet{Giallongo_2019}, while we caution that neither of them are spectroscopically confirmed.

\section{Future Prospects}\label{sec:summary}
Spectroscopic follow-up observations are the key to confirming the redshift and the nature of \target, and also our estimate of the $z=5$ AGN luminosity function at the very faint end, where galaxies are dominant against AGNs in number (Figure~\ref{fig:qlf}).
Those observations will also help to test the extreme emission-line properties that we discuss in Sections~\ref{sec:result_BBexcess} and \ref{sec:result_sedfit}.
Specifically, detection of broad-line emission will enable us to perform a virial BH mass estimate via Balmer lines or other mass tracers such as Mg~{\sc ii}, from which one can constrain the mass distribution of seed BHs in the earlier epochs of the universe (\citealt{Toyouchi_2022}; see also \citealt{Inayoshi20, Greene_2020}).

Our initial result based on the first observations of the CEERS program indicates that deep NIR imaging observations of \JWST\ are capable of determining the high-redshift AGN luminosity function at the very faint end.
More complete and sophisticated AGN selection criteria are required to better constrain the underlying AGN population.
Ongoing and upcoming wide-field surveys with \JWST\ such as COSMOS Web and JADES programs can also be used to perform wider and deeper surveys of high-redshift AGNs.

\begin{acknowledgments}

We wish to thank the entire \JWST\ team and the CEERS collaboration for the operation of the telescope and for developing their observing program with a zero-exclusive-access period.
This work is based  on observations made with the NASA/ESA/CSA James Webb Space Telescope. The data were obtained from the Mikulski Archive for Space Telescopes at the Space Telescope Science Institute, which is operated by the Association of Universities for Research in Astronomy, Inc., under NASA contract NAS5-03127 and NAS5–26555 for JWST. These observations are associated with program \#1345.
The specific observations analyzed can be accessed via \dataset[https://doi.org/10.17909/3pf0-8b20]{https://doi.org/10.17909/3pf0-8b20}. 
Support to MAST for these data is provided by the NASA Office of Space Science via grant NAG5–7584 and by other grants and contracts.

This work has made use of data from the European Space Agency (ESA) mission
{\it Gaia} (\url{https://www.cosmos.esa.int/gaia}), processed by the {\it Gaia}
Data Processing and Analysis Consortium (DPAC,
\url{https://www.cosmos.esa.int/web/gaia/dpac/consortium}). Funding for the DPAC
has been provided by national institutions, in particular the institutions
participating in the {\it Gaia} Multilateral Agreement.

We acknowledge support from the National Natural Science Foundation of China (12150410307, 12073003, 12003003, 11721303, 11991052, 11950410493), and the China Manned Space Project Nos. CMS-CSST-2021-A04 and CMS-CSST-2021-A06. 
X. D. is supported by JSPS KAKENHI Grant Number JP22K14071.

\end{acknowledgments}

\vspace{5mm}
\facilities{\JWST, \HST, \Spitzer, \CFHT, GAIA}

\software{astropy \citep{Astropy},  
          \textsf{Galight} \citep{Ding20, Ding21},
          \textsf{Photutils} \citep{Photutils_v140},
          \textsf{SExtractor} \citep{SExtractor},
          STScI JWST Calibration Pipeline (\url{jwst-pipeline.readthedocs.io})
          }

\bibliography{ref}{}

\begin{thebibliography}{}
\expandafter\ifx\csname natexlab\endcsname\relax\def\natexlab#1{#1}\fi
\providecommand{\url}[1]{\href{#1}{#1}}
\providecommand{\dodoi}[1]{doi:~\href{http://doi.org/#1}{\nolinkurl{#1}}}
\providecommand{\doeprint}[1]{\href{http://ascl.net/#1}{\nolinkurl{http://ascl.net/#1}}}
\providecommand{\doarXiv}[1]{\href{https://arxiv.org/abs/#1}{\nolinkurl{https://arxiv.org/abs/#1}}}

\bibitem[{{Adams} {et~al.}(2022{\natexlab{a}}){Adams}, {Bowler}, {Jarvis},
  {Varadaraj}, \& {H{\"a}u{\ss}ler}}]{Adams22a}
{Adams}, N.~J., {Bowler}, R.~A.~A., {Jarvis}, M.~J., {Varadaraj}, R.~G., \&
  {H{\"a}u{\ss}ler}, B. 2022{\natexlab{a}}, arXiv e-prints, arXiv:2207.09342.
\newblock \doarXiv{2207.09342}

\bibitem[{{Adams} {et~al.}(2022{\natexlab{b}}){Adams}, {Conselice}, {Ferreira},
  {Austin}, {Trussler}, {Juod{\v{z}}balis}, {Wilkins}, {Caruana}, {Dayal},
  {Verma}, \& {Vijayan}}]{Adams22b}
{Adams}, N.~J., {Conselice}, C.~J., {Ferreira}, L., {et~al.}
  2022{\natexlab{b}}, arXiv e-prints, arXiv:2207.11217.
\newblock \doarXiv{2207.11217}

\bibitem[{{Astropy Collaboration} {et~al.}(2013){Astropy Collaboration},
  {Robitaille}, {Tollerud}, {Greenfield}, {Droettboom}, {Bray}, {Aldcroft},
  {Davis}, {Ginsburg}, {Price-Whelan}, {Kerzendorf}, {Conley}, {Crighton},
  {Barbary}, {Muna}, {Ferguson}, {Grollier}, {Parikh}, {Nair}, {Unther},
  {Deil}, {Woillez}, {Conseil}, {Kramer}, {Turner}, {Singer}, {Fox}, {Weaver},
  {Zabalza}, {Edwards}, {Azalee Bostroem}, {Burke}, {Casey}, {Crawford},
  {Dencheva}, {Ely}, {Jenness}, {Labrie}, {Lim}, {Pierfederici}, {Pontzen},
  {Ptak}, {Refsdal}, {Servillat}, \& {Streicher}}]{Astropy}
{Astropy Collaboration}, {Robitaille}, T.~P., {Tollerud}, E.~J., {et~al.} 2013,
  \aap, 558, A33, \dodoi{10.1051/0004-6361/201322068}

\bibitem[{{Ba{\~n}ados} {et~al.}(2018){Ba{\~n}ados}, {Venemans},
  {Mazzucchelli}, {Farina}, {Walter}, {Wang}, {Decarli}, {Stern}, {Fan},
  {Davies}, {Hennawi}, {Simcoe}, {Turner}, {Rix}, {Yang}, {Kelson}, {Rudie}, \&
  {Winters}}]{Banados18}
{Ba{\~n}ados}, E., {Venemans}, B.~P., {Mazzucchelli}, C., {et~al.} 2018, \nat,
  553, 473, \dodoi{10.1038/nature25180}

\bibitem[{{Bagley} {et~al.}(2022){Bagley}, {Finkelstein}, {Koekemoer},
  {Ferguson}, {Arrabal Haro}, {Dickinson}, {Kartaltepe}, {Papovich},
  {P{\'e}rez-Gonz{\'a}lez}, {Pirzkal}, {Somerville}, {Willmer}, {Yang}, {Yung},
  {Fontana}, {Grazian}, {Grogin}, {Hirschmann}, {Kewley}, {Kirkpatrick},
  {Kocevski}, {Lotz}, {Medrano}, {Morales}, {Pentericci}, {Ravindranath},
  {Trump}, {Wilkins}, {Calabr{\`o}}, {Cooper}, {Costantin}, {de la Vega},
  {Hutchison}, {Lucas}, {McGrath}, {Wang}, \& {Wuyts}}]{Bagley22}
{Bagley}, M.~B., {Finkelstein}, S.~L., {Koekemoer}, A.~M., {et~al.} 2022, arXiv
  e-prints, arXiv:2211.02495.
\newblock \doarXiv{2211.02495}

\bibitem[{{Baldwin}(1977)}]{Baldwin77}
{Baldwin}, J.~A. 1977, \apj, 214, 679, \dodoi{10.1086/155294}

\bibitem[{{Bertin} \& {Arnouts}(1996)}]{SExtractor}
{Bertin}, E., \& {Arnouts}, S. 1996, \aaps, 117, 393,
  \dodoi{10.1051/aas:1996164}

\bibitem[{{Bouwens} {et~al.}(2021){Bouwens}, {Oesch}, {Stefanon},
  {Illingworth}, {Labb{\'e}}, {Reddy}, {Atek}, {Montes}, {Naidu},
  {Nanayakkara}, {Nelson}, \& {Wilkins}}]{Bouwens21}
{Bouwens}, R.~J., {Oesch}, P.~A., {Stefanon}, M., {et~al.} 2021, \aj, 162, 47,
  \dodoi{10.3847/1538-3881/abf83e}

\bibitem[{{Bowler} {et~al.}(2021){Bowler}, {Adams}, {Jarvis}, \&
  {H{\"a}u{\ss}ler}}]{Bowler21}
{Bowler}, R.~A.~A., {Adams}, N.~J., {Jarvis}, M.~J., \& {H{\"a}u{\ss}ler}, B.
  2021, \mnras, 502, 662, \dodoi{10.1093/mnras/stab038}

\bibitem[{{Bradley} {et~al.}(2022){Bradley}, {Sip{\H{o}}cz}, {Robitaille},
  {Tollerud}, {Vin{\'\i}cius}, {Deil}, {Barbary}, {Wilson}, {Busko}, {Donath},
  {G{\"u}nther}, {Cara}, {Lim}, {Me{\ss}linger}, {Conseil}, {Bostroem},
  {Droettboom}, {Bray}, {Andersen Bratholm}, {Barentsen}, {Craig}, {Rathi},
  {Pascual}, {Perren}, {Georgiev}, {De Val-Borro}, {Kerzendorf}, {Bach},
  {Quint}, \& {Souchereau}}]{Photutils_v140}
{Bradley}, L., {Sip{\H{o}}cz}, B., {Robitaille}, T., {et~al.} 2022,
  {astropy/photutils:}, 1.4.0, Zenodo,  Zenodo, \dodoi{10.5281/zenodo.6385735}

\bibitem[{{Calzetti} {et~al.}(2000){Calzetti}, {Armus}, {Bohlin}, {Kinney},
  {Koornneef}, \& {Storchi-Bergmann}}]{Calzetti00}
{Calzetti}, D., {Armus}, L., {Bohlin}, R.~C., {et~al.} 2000, \apj, 533, 682,
  \dodoi{10.1086/308692}

\bibitem[{{Castellano} {et~al.}(2022){Castellano}, {Fontana}, {Treu},
  {Santini}, {Merlin}, {Leethochawalit}, {Trenti}, {Mestric}, {Vanzella},
  {Bonchi}, {Belfiori}, {Nonino}, {Paris}, {Polenta}, {Roberts-Borsani},
  {Boyett}, {Calabro}, {Glazebrook}, {Grillo}, {Mascia}, {Mason}, {Mercurio},
  {Morishita}, {Nanayakkara}, {Pentericci}, {Rosati}, {Vulcani}, {Wang}, \&
  {Yang}}]{Castellano_2022}
{Castellano}, M., {Fontana}, A., {Treu}, T., {et~al.} 2022, arXiv e-prints,
  arXiv:2207.09436.
\newblock \doarXiv{2207.09436}

\bibitem[{{Chen} {et~al.}(2022){Chen}, {Stark}, {Endsley}, {Topping},
  {Whitler}, \& {Charlot}}]{Chen22}
{Chen}, Z., {Stark}, D.~P., {Endsley}, R., {et~al.} 2022, arXiv e-prints,
  arXiv:2207.12657.
\newblock \doarXiv{2207.12657}

\bibitem[{{Ding} {et~al.}(2021){Ding}, {Birrer}, {Treu}, \&
  {Silverman}}]{Ding21}
{Ding}, X., {Birrer}, S., {Treu}, T., \& {Silverman}, J.~D. 2021, arXiv
  e-prints, arXiv:2111.08721.
\newblock \doarXiv{2111.08721}

\bibitem[{{Ding} {et~al.}(2022){Ding}, {Silverman}, \& {Onoue}}]{Ding22}
{Ding}, X., {Silverman}, J.~D., \& {Onoue}, M. 2022, arXiv e-prints,
  arXiv:2209.03359.
\newblock \doarXiv{2209.03359}

\bibitem[{{Ding} {et~al.}(2020){Ding}, {Silverman}, {Treu}, {Schulze},
  {Schramm}, {Birrer}, {Park}, {Jahnke}, {Bennert}, {Kartaltepe}, {Koekemoer},
  {Malkan}, \& {Sanders}}]{Ding20}
{Ding}, X., {Silverman}, J., {Treu}, T., {et~al.} 2020, \apj, 888, 37,
  \dodoi{10.3847/1538-4357/ab5b90}

\bibitem[{{Donnan} {et~al.}(2022){Donnan}, {McLeod}, {Dunlop}, {McLure},
  {Carnall}, {Begley}, {Cullen}, {Hamadouche}, {Bowler}, {McCracken},
  {Milvang-Jensen}, {Moneti}, \& {Targett}}]{Donnan_2022}
{Donnan}, C.~T., {McLeod}, D.~J., {Dunlop}, J.~S., {et~al.} 2022, arXiv
  e-prints, arXiv:2207.12356.
\newblock \doarXiv{2207.12356}

\bibitem[{{Endsley} {et~al.}(2021){Endsley}, {Stark}, {Chevallard}, \&
  {Charlot}}]{Endsley21}
{Endsley}, R., {Stark}, D.~P., {Chevallard}, J., \& {Charlot}, S. 2021, \mnras,
  500, 5229, \dodoi{10.1093/mnras/staa3370}

\bibitem[{{Endsley} {et~al.}(2022){Endsley}, {Stark}, {Whitler}, {Topping},
  {Chen}, {Plat}, {Chisholm}, \& {Charlot}}]{Endsley22}
{Endsley}, R., {Stark}, D.~P., {Whitler}, L., {et~al.} 2022, arXiv e-prints,
  arXiv:2208.14999.
\newblock \doarXiv{2208.14999}

\bibitem[{{Fan} {et~al.}(2001{\natexlab{a}}){Fan}, {Narayanan}, {Lupton},
  {Strauss}, {Knapp}, {Becker}, {White}, {Pentericci}, {Leggett}, {Haiman},
  {Gunn}, {Ivezi{\'c}}, {Schneider}, {Anderson}, {Brinkmann}, {Bahcall},
  {Connolly}, {Csabai}, {Doi}, {Fukugita}, {Geballe}, {Grebel}, {Harbeck},
  {Hennessy}, {Lamb}, {Miknaitis}, {Munn}, {Nichol}, {Okamura}, {Pier},
  {Prada}, {Richards}, {Szalay}, \& {York}}]{Fan01}
{Fan}, X., {Narayanan}, V.~K., {Lupton}, R.~H., {et~al.} 2001{\natexlab{a}},
  \aj, 122, 2833, \dodoi{10.1086/324111}

\bibitem[{{Fan} {et~al.}(2001{\natexlab{b}}){Fan}, {Strauss}, {Richards},
  {Newman}, {Becker}, {Schneider}, {Gunn}, {Davis}, {White}, {Lupton},
  {Anderson}, {Annis}, {Bahcall}, {Brunner}, {Csabai}, {Doi}, {Fukugita},
  {Hennessy}, {Hindsley}, {Ivezi{\'c}}, {Knapp}, {McKay}, {Munn}, {Pier},
  {Szalay}, \& {York}}]{Fan01c}
{Fan}, X., {Strauss}, M.~A., {Richards}, G.~T., {et~al.} 2001{\natexlab{b}},
  \aj, 121, 31, \dodoi{10.1086/318032}

\bibitem[{{Ferland} {et~al.}(2017){Ferland}, {Chatzikos}, {Guzm{\'a}n},
  {Lykins}, {van Hoof}, {Williams}, {Abel}, {Badnell}, {Keenan}, {Porter}, \&
  {Stancil}}]{Ferland_2017}
{Ferland}, G.~J., {Chatzikos}, M., {Guzm{\'a}n}, F., {et~al.} 2017, \rmxaa, 53,
  385.
\newblock \doarXiv{1705.10877}

\bibitem[{{Finkelstein} \& {Bagley}(2022)}]{Finkelstein22b}
{Finkelstein}, S.~L., \& {Bagley}, M.~B. 2022, arXiv e-prints,
  arXiv:2207.02233.
\newblock \doarXiv{2207.02233}

\bibitem[{{Finkelstein} {et~al.}(2017){Finkelstein}, {Dickinson}, {Ferguson},
  {Grazian}, {Grogin}, {Kartaltepe}, {Kewley}, {Kocevski}, {Koekemoer}, {Lotz},
  {Papovich}, {Pentericci}, {Perez-Gonzalez}, {Pirzkal}, {Ravindranath},
  {Somerville}, {Trump}, \& {Wilkins}}]{CEERS}
{Finkelstein}, S.~L., {Dickinson}, M., {Ferguson}, H.~C., {et~al.} 2017, {The
  Cosmic Evolution Early Release Science (CEERS) Survey}, JWST Proposal ID
  1345. Cycle 0 Early Release Science

\bibitem[{{Finkelstein} {et~al.}(2022{\natexlab{a}}){Finkelstein}, {Bagley},
  {Arrabal Haro}, {Dickinson}, {Ferguson}, {Kartaltepe}, {Papovich},
  {Burgarella}, {Kocevski}, {Huertas-Company}, {Iyer}, {Larson},
  {P{\'e}rez-Gonz{\'a}lez}, {Rose}, {Tacchella}, {Wilkins}, {Chworowsky},
  {Medrano}, {Morales}, {Somerville}, {Yung}, {Fontana}, {Giavalisco},
  {Grazian}, {Grogin}, {Kewley}, {Koekemoer}, {Kirkpatrick}, {Kurczynski},
  {Lotz}, {Pentericci}, {Pirzkal}, {Ravindranath}, {Ryan}, {Trump}, {Yang},
  {Almaini}, {Amor{\'\i}n}, {Annunziatella}, {Backhaus}, {Barro}, {Behroozi},
  {Bell}, {Bhatawdekar}, {Bisigello}, {Bromm}, {Buat}, {Buitrago},
  {Calabr{\'o}}, {Casey}, {Castellano}, {Ch{\'a}vez Ortiz}, {Ciesla}, {Cleri},
  {Cohen}, {Cole}, {Cooke}, {Cooper}, {Cooray}, {Costantin}, {Cox}, {Croton},
  {Daddi}, {Dav{\'e}}, {de la Vega}, {Dekel}, {Elbaz}, {Estrada-Carpenter},
  {Faber}, {Fern{\'a}ndez}, {Finkelstein}, {Freundlich}, {Fujimoto},
  {Garc{\'\i}a-Argum{\'a}nez}, {Gardner}, {Gawiser}, {G{\'o}mez-Guijarro},
  {Guo}, {Hamilton}, {Hathi}, {Holwerda}, {Hirschmann}, {Hutchison}, {Jha},
  {Jogee}, {Juneau}, {Jung}, {Kassin}, {Le Bail}, {Leung}, {Lucas}, {Magnelli},
  {Mantha}, {Matharu}, {McGrath}, {McIntosh}, {Merlin}, {Mobasher}, {Newman},
  {Nicholls}, {Pandya}, {Rafelski}, {Ronayne}, {Santini}, {Seill{\'e}}, {Shah},
  {Shen}, {Simons}, {Snyder}, {Stanway}, {Straughn}, {Teplitz}, {Vanderhoof},
  {Vega-Ferrero}, {Wang}, {Weiner}, {Willmer}, {Wuyts}, \&
  {Zavala}}]{Finkelstein22}
{Finkelstein}, S.~L., {Bagley}, M.~B., {Arrabal Haro}, P., {et~al.}
  2022{\natexlab{a}}, arXiv e-prints, arXiv:2207.12474.
\newblock \doarXiv{2207.12474}

\bibitem[{{Finkelstein} {et~al.}(2022{\natexlab{b}}){Finkelstein}, {Bagley},
  {Ferguson}, {Wilkins}, {Kartaltepe}, {Papovich}, {Yung}, {Arrabal Haro},
  {Behroozi}, {Dickinson}, {Kocevski}, {Koekemoer}, {Larson}, {Le Bail},
  {Morales}, {Perez-Gonzalez}, {Burgarella}, {Dave}, {Hirschmann},
  {Somerville}, {Wuyts}, {Bromm}, {Casey}, {Fontana}, {Fujimoto}, {Gardner},
  {Giavalisco}, {Grazian}, {Grogin}, {Hathi}, {Hutchison}, {Jha}, {Jogee},
  {Kewley}, {Kirkpatrick}, {Long}, {Lotz}, {Pentericci}, {Pierel}, {Pirzkal},
  {Ravindranath}, {Ryan}, {Trump}, {Yang}, {Bhatawdekar}, {Bisigello}, {Buat},
  {Calabro}, {Castellano}, {Cleri}, {Cooper}, {Croton}, {Daddi}, {Dekel},
  {Elbaz}, {Franco}, {Gawiser}, {Holwerda}, {Huertas-Company}, {Jaskot},
  {Leung}, {Lucas}, {Mobasher}, {Pandya}, {Tacchella}, {Weiner}, \&
  {Zavala}}]{Finkelstein22c}
{Finkelstein}, S.~L., {Bagley}, M.~B., {Ferguson}, H.~C., {et~al.}
  2022{\natexlab{b}}, arXiv e-prints, arXiv:2211.05792.
\newblock \doarXiv{2211.05792}

\bibitem[{{Fujimoto} {et~al.}(2022){Fujimoto}, {Brammer}, {Watson}, {Magdis},
  {Kokorev}, {Greve}, {Toft}, {Walter}, {Valiante}, {Ginolfi}, {Schneider},
  {Valentino}, {Colina}, {Vestergaard}, {Marques-Chaves}, {Fynbo}, {Krips},
  {Steinhardt}, {Cortzen}, {Rizzo}, \& {Oesch}}]{Fujimoto22}
{Fujimoto}, S., {Brammer}, G.~B., {Watson}, D., {et~al.} 2022, \nat, 604, 261,
  \dodoi{10.1038/s41586-022-04454-1}

\bibitem[{{Gaia Collaboration} {et~al.}(2016){Gaia Collaboration}, {Prusti},
  {de Bruijne}, {Brown}, {Vallenari}, {Babusiaux}, {Bailer-Jones}, {Bastian},
  {Biermann}, {Evans}, \& et~al.}]{GAIA16}
{Gaia Collaboration}, {Prusti}, T., {de Bruijne}, J.~H.~J., {et~al.} 2016,
  \aap, 595, A1, \dodoi{10.1051/0004-6361/201629272}

\bibitem[{{Giallongo} {et~al.}(2019){Giallongo}, {Grazian}, {Fiore}, {Kodra},
  {Urrutia}, {Castellano}, {Cristiani}, {Dickinson}, {Fontana}, {Menci},
  {Pentericci}, {Boutsia}, {Newman}, \& {Puccetti}}]{Giallongo_2019}
{Giallongo}, E., {Grazian}, A., {Fiore}, F., {et~al.} 2019, \apj, 884, 19,
  \dodoi{10.3847/1538-4357/ab39e1}

\bibitem[{{Goulding} \& {Greene}(2022)}]{Goulding22}
{Goulding}, A.~D., \& {Greene}, J.~E. 2022, arXiv e-prints, arXiv:2208.02822.
\newblock \doarXiv{2208.02822}

\bibitem[{{Greene} \& {Ho}(2005)}]{GH05}
{Greene}, J.~E., \& {Ho}, L.~C. 2005, \apj, 630, 122, \dodoi{10.1086/431897}

\bibitem[{{Greene} \& {Ho}(2007)}]{GH07c}
---. 2007, \apj, 670, 92, \dodoi{10.1086/522082}

\bibitem[{{Greene} {et~al.}(2020){Greene}, {Strader}, \& {Ho}}]{Greene_2020}
{Greene}, J.~E., {Strader}, J., \& {Ho}, L.~C. 2020, \araa, 58, 257,
  \dodoi{10.1146/annurev-astro-032620-021835}

\bibitem[{{Harikane} {et~al.}(2022{\natexlab{a}}){Harikane}, {Ono}, {Ouchi},
  {Liu}, {Sawicki}, {Shibuya}, {Behroozi}, {He}, {Shimasaku}, {Arnouts},
  {Coupon}, {Fujimoto}, {Gwyn}, {Huang}, {Inoue}, {Kashikawa}, {Komiyama},
  {Matsuoka}, \& {Willott}}]{Harikane22a}
{Harikane}, Y., {Ono}, Y., {Ouchi}, M., {et~al.} 2022{\natexlab{a}}, \apjs,
  259, 20, \dodoi{10.3847/1538-4365/ac3dfc}

\bibitem[{{Harikane} {et~al.}(2022{\natexlab{b}}){Harikane}, {Ouchi}, {Oguri},
  {Ono}, {Nakajima}, {Isobe}, {Umeda}, {Mawatari}, \& {Zhang}}]{Harikane_2022c}
{Harikane}, Y., {Ouchi}, M., {Oguri}, M., {et~al.} 2022{\natexlab{b}}, arXiv
  e-prints, arXiv:2208.01612.
\newblock \doarXiv{2208.01612}

\bibitem[{{Heckman} {et~al.}(2004){Heckman}, {Kauffmann}, {Brinchmann},
  {Charlot}, {Tremonti}, \& {White}}]{Heckman_2004}
{Heckman}, T.~M., {Kauffmann}, G., {Brinchmann}, J., {et~al.} 2004, \apj, 613,
  109, \dodoi{10.1086/422872}

\bibitem[{{Hu} {et~al.}(2022){Hu}, {Inayoshi}, {Haiman}, {Li}, {Quataert}, \&
  {Kuiper}}]{Hu_2022b}
{Hu}, H., {Inayoshi}, K., {Haiman}, Z., {et~al.} 2022, \apj, 935, 140,
  \dodoi{10.3847/1538-4357/ac7daa}

\bibitem[{{Inayoshi} {et~al.}(2022{\natexlab{a}}){Inayoshi}, {Nakatani},
  {Toyouchi}, {Hosokawa}, {Kuiper}, \& {Onoue}}]{Inayoshi_2022a}
{Inayoshi}, K., {Nakatani}, R., {Toyouchi}, D., {et~al.} 2022{\natexlab{a}},
  \apj, 927, 237, \dodoi{10.3847/1538-4357/ac4751}

\bibitem[{{Inayoshi} {et~al.}(2022{\natexlab{b}}){Inayoshi}, {Onoue},
  {Sugahara}, {Inoue}, \& {Ho}}]{Inayoshi22}
{Inayoshi}, K., {Onoue}, M., {Sugahara}, Y., {Inoue}, A.~K., \& {Ho}, L.~C.
  2022{\natexlab{b}}, \apjl, 931, L25, \dodoi{10.3847/2041-8213/ac6f01}

\bibitem[{{Inayoshi} {et~al.}(2020){Inayoshi}, {Visbal}, \&
  {Haiman}}]{Inayoshi20}
{Inayoshi}, K., {Visbal}, E., \& {Haiman}, Z. 2020, \araa, 58, 27,
  \dodoi{10.1146/annurev-astro-120419-014455}

\bibitem[{{Inoue}(2011)}]{Inoue11}
{Inoue}, A.~K. 2011, \mnras, 415, 2920,
  \dodoi{10.1111/j.1365-2966.2011.18906.x}

\bibitem[{{Inoue} {et~al.}(2020){Inoue}, {Hashimoto}, {Chihara}, \&
  {Koike}}]{Inoue20}
{Inoue}, A.~K., {Hashimoto}, T., {Chihara}, H., \& {Koike}, C. 2020, \mnras,
  495, 1577, \dodoi{10.1093/mnras/staa1203}

\bibitem[{{Ishikawa} {et~al.}(2022){Ishikawa}, {Morishita}, {Stiavelli},
  {Leethochawalit}, {Ferguson}, {Gilli}, {Mason}, {Trenti}, {Treu}, \&
  {Norman}}]{Ishikawa22}
{Ishikawa}, Y., {Morishita}, T., {Stiavelli}, M., {et~al.} 2022, \apj, 936,
  167, \dodoi{10.3847/1538-4357/ac8874}

\bibitem[{{Jiang} {et~al.}(2016){Jiang}, {McGreer}, {Fan}, {Strauss},
  {Ba{\~n}ados}, {Becker}, {Bian}, {Farnsworth}, {Shen}, {Wang}, {Wang},
  {Wang}, {White}, {Wu}, {Wu}, {Yang}, \& {Yang}}]{Jiang16}
{Jiang}, L., {McGreer}, I.~D., {Fan}, X., {et~al.} 2016, \apj, 833, 222,
  \dodoi{10.3847/1538-4357/833/2/222}

\bibitem[{{Jiang} {et~al.}(2022){Jiang}, {Ning}, {Fan}, {Ho}, {Luo}, {Wang},
  {Wu}, {Wu}, {Yang}, \& {Zheng}}]{Jiang22}
{Jiang}, L., {Ning}, Y., {Fan}, X., {et~al.} 2022, Nature Astronomy, 6, 850,
  \dodoi{10.1038/s41550-022-01708-w}

\bibitem[{{Kato} {et~al.}(2020){Kato}, {Matsuoka}, {Onoue}, {Koyama}, {Toba},
  {Akiyama}, {Fujimoto}, {Imanishi}, {Iwasawa}, {Izumi}, {Kashikawa},
  {Kawaguchi}, {Lee}, {Minezaki}, {Nagao}, {Noboriguchi}, \&
  {Strauss}}]{Kato20}
{Kato}, N., {Matsuoka}, Y., {Onoue}, M., {et~al.} 2020, \pasj, 72, 84,
  \dodoi{10.1093/pasj/psaa074}

\bibitem[{{Kong} \& {Ho}(2018)}]{KH18}
{Kong}, M., \& {Ho}, L.~C. 2018, \apj, 859, 116,
  \dodoi{10.3847/1538-4357/aabe2a}

\bibitem[{{Laporte} {et~al.}(2017){Laporte}, {Nakajima}, {Ellis}, {Zitrin},
  {Stark}, {Mainali}, \& {Roberts-Borsani}}]{Laporte17}
{Laporte}, N., {Nakajima}, K., {Ellis}, R.~S., {et~al.} 2017, \apj, 851, 40,
  \dodoi{10.3847/1538-4357/aa96a8}

\bibitem[{{Li} {et~al.}(2022){Li}, {Inayoshi}, {Onoue}, \& {Toyouchi}}]{Li22}
{Li}, W., {Inayoshi}, K., {Onoue}, M., \& {Toyouchi}, D. 2022, arXiv e-prints,
  arXiv:2210.02308.
\newblock \doarXiv{2210.02308}

\bibitem[{{Liu} {et~al.}(2019){Liu}, {Liu}, {Dong}, {Zhou}, {Wang}, {Lu}, \&
  {Yuan}}]{Liu19}
{Liu}, H.-Y., {Liu}, W.-J., {Dong}, X.-B., {et~al.} 2019, \apjs, 243, 21,
  \dodoi{10.3847/1538-4365/ab298b}

\bibitem[{{Liu} {et~al.}(2018){Liu}, {Yuan}, {Dong}, {Zhou}, \& {Liu}}]{Liu18}
{Liu}, H.-Y., {Yuan}, W., {Dong}, X.-B., {Zhou}, H., \& {Liu}, W.-J. 2018,
  \apjs, 235, 40, \dodoi{10.3847/1538-4365/aab88e}

\bibitem[{{Madau}(1995)}]{Madau95}
{Madau}, P. 1995, \apj, 441, 18, \dodoi{10.1086/175332}

\bibitem[{{Matsuoka} {et~al.}(2018){Matsuoka}, {Strauss}, {Kashikawa}, {Onoue},
  {Iwasawa}, {Tang}, {Lee}, {Imanishi}, {Nagao}, {Akiyama}, {Asami}, {Bosch},
  {Furusawa}, {Goto}, {Gunn}, {Harikane}, {Ikeda}, {Izumi}, {Kawaguchi},
  {Kato}, {Kikuta}, {Kohno}, {Komiyama}, {Lupton}, {Minezaki}, {Miyazaki},
  {Murayama}, {Niida}, {Nishizawa}, {Noboriguchi}, {Oguri}, {Ono}, {Ouchi},
  {Price}, {Sameshima}, {Schulze}, {Shirakata}, {Silverman}, {Sugiyama},
  {Tait}, {Takada}, {Takata}, {Tanaka}, {Toba}, {Utsumi}, {Wang}, \&
  {Yamashita}}]{Matsuoka18c}
{Matsuoka}, Y., {Strauss}, M.~A., {Kashikawa}, N., {et~al.} 2018, \apj, 869,
  150, \dodoi{10.3847/1538-4357/aaee7a}

\bibitem[{{Matsuoka} {et~al.}(2019){Matsuoka}, {Onoue}, {Kashikawa}, {Strauss},
  {Iwasawa}, {Lee}, {Imanishi}, {Nagao}, {Akiyama}, {Asami}, {Bosch},
  {Furusawa}, {Goto}, {Gunn}, {Harikane}, {Ikeda}, {Izumi}, {Kawaguchi},
  {Kato}, {Kikuta}, {Kohno}, {Komiyama}, {Koyama}, {Lupton}, {Minezaki},
  {Miyazaki}, {Murayama}, {Niida}, {Nishizawa}, {Noboriguchi}, {Oguri}, {Ono},
  {Ouchi}, {Price}, {Sameshima}, {Schulze}, {Shirakata}, {Silverman},
  {Sugiyama}, {Tait}, {Takada}, {Takata}, {Tanaka}, {Tang}, {Toba}, {Utsumi},
  {Wang}, \& {Yamashita}}]{Matsuoka19a}
{Matsuoka}, Y., {Onoue}, M., {Kashikawa}, N., {et~al.} 2019, \apjl, 872, L2,
  \dodoi{10.3847/2041-8213/ab0216}

\bibitem[{{McGreer} {et~al.}(2018){McGreer}, {Fan}, {Jiang}, \&
  {Cai}}]{McGreer_2018}
{McGreer}, I.~D., {Fan}, X., {Jiang}, L., \& {Cai}, Z. 2018, \aj, 155, 131,
  \dodoi{10.3847/1538-3881/aaaab4}

\bibitem[{{Morishita} {et~al.}(2020){Morishita}, {Stiavelli}, {Trenti}, {Treu},
  {Roberts-Borsani}, {Mason}, {Hashimoto}, {Bradley}, {Coe}, \&
  {Ishikawa}}]{Morishita20}
{Morishita}, T., {Stiavelli}, M., {Trenti}, M., {et~al.} 2020, \apj, 904, 50,
  \dodoi{10.3847/1538-4357/abba83}

\bibitem[{{Naidu} {et~al.}(2022){Naidu}, {Oesch}, {Setton}, {Matthee},
  {Conroy}, {Johnson}, {Weaver}, {Bouwens}, {Brammer}, {Dayal}, {Illingworth},
  {Barrufet}, {Belli}, {Bezanson}, {Bose}, {Heintz}, {Leja}, {Leonova},
  {Marques-Chaves}, {Stefanon}, {Toft}, {van der Wel}, {van Dokkum}, {Weibel},
  \& {Whitaker}}]{Naidu_2022b}
{Naidu}, R.~P., {Oesch}, P.~A., {Setton}, D.~J., {et~al.} 2022, arXiv e-prints,
  arXiv:2208.02794.
\newblock \doarXiv{2208.02794}

\bibitem[{{Nandra} {et~al.}(2015){Nandra}, {Laird}, {Aird}, {Salvato},
  {Georgakakis}, {Barro}, {Perez-Gonzalez}, {Barmby}, {Chary}, {Coil},
  {Cooper}, {Davis}, {Dickinson}, {Faber}, {Fazio}, {Guhathakurta}, {Gwyn},
  {Hsu}, {Huang}, {Ivison}, {Koo}, {Newman}, {Rangel}, {Yamada}, \&
  {Willmer}}]{Nandra15}
{Nandra}, K., {Laird}, E.~S., {Aird}, J.~A., {et~al.} 2015, \apjs, 220, 10,
  \dodoi{10.1088/0067-0049/220/1/10}

\bibitem[{{Natarajan} {et~al.}(2017){Natarajan}, {Pacucci}, {Ferrara},
  {Agarwal}, {Ricarte}, {Zackrisson}, \& {Cappelluti}}]{Natarajan_2017}
{Natarajan}, P., {Pacucci}, F., {Ferrara}, A., {et~al.} 2017, \apj, 838, 117,
  \dodoi{10.3847/1538-4357/aa6330}

\bibitem[{{Neeleman} {et~al.}(2021){Neeleman}, {Novak}, {Venemans}, {Walter},
  {Decarli}, {Kaasinen}, {Schindler}, {Ba{\~n}ados}, {Carilli}, {Drake}, {Fan},
  \& {Rix}}]{Neeleman21}
{Neeleman}, M., {Novak}, M., {Venemans}, B.~P., {et~al.} 2021, \apj, 911, 141,
  \dodoi{10.3847/1538-4357/abe70f}

\bibitem[{{Netzer}(2019)}]{Netzer19}
{Netzer}, H. 2019, \mnras, 488, 5185, \dodoi{10.1093/mnras/stz2016}

\bibitem[{{Niida} {et~al.}(2020){Niida}, {Nagao}, {Ikeda}, {Akiyama},
  {Matsuoka}, {He}, {Matsuoka}, {Toba}, {Onoue}, {Kobayashi}, {Taniguchi},
  {Furusawa}, {Harikane}, {Imanishi}, {Kashikawa}, {Kawaguchi}, {Komiyama},
  {Shirakata}, {Terashima}, \& {Ueda}}]{Niida20}
{Niida}, M., {Nagao}, T., {Ikeda}, H., {et~al.} 2020, \apj, 904, 89,
  \dodoi{10.3847/1538-4357/abbe11}

\bibitem[{{Ono} {et~al.}(2018){Ono}, {Ouchi}, {Harikane}, {Toshikawa}, {Rauch},
  {Yuma}, {Sawicki}, {Shibuya}, {Shimasaku}, {Oguri}, {Willott}, {Akhlaghi},
  {Akiyama}, {Coupon}, {Kashikawa}, {Komiyama}, {Konno}, {Lin}, {Matsuoka},
  {Miyazaki}, {Nagao}, {Nakajima}, {Silverman}, {Tanaka}, {Taniguchi}, \&
  {Wang}}]{Ono18}
{Ono}, Y., {Ouchi}, M., {Harikane}, Y., {et~al.} 2018, \pasj, 70, S10,
  \dodoi{10.1093/pasj/psx103}

\bibitem[{{Ono} {et~al.}(2022){Ono}, {Harikane}, {Ouchi}, {Yajima}, {Abe},
  {Isobe}, {Shibuya}, {Zhang}, {Nakajima}, \& {Umeda}}]{Ono22}
{Ono}, Y., {Harikane}, Y., {Ouchi}, M., {et~al.} 2022, arXiv e-prints,
  arXiv:2208.13582.
\newblock \doarXiv{2208.13582}

\bibitem[{{Onoue} {et~al.}(2019){Onoue}, {Kashikawa}, {Matsuoka}, {Kato},
  {Izumi}, {Nagao}, {Strauss}, {Harikane}, {Imanishi}, {Ito}, {Iwasawa},
  {Kawaguchi}, {Lee}, {Noboriguchi}, {Suh}, {Tanaka}, \& {Toba}}]{Onoue19}
{Onoue}, M., {Kashikawa}, N., {Matsuoka}, Y., {et~al.} 2019, \apj, 880, 77,
  \dodoi{10.3847/1538-4357/ab29e9}

\bibitem[{{Prevot} {et~al.}(1984){Prevot}, {Lequeux}, {Maurice}, {Prevot}, \&
  {Rocca-Volmerange}}]{Prevot_1984}
{Prevot}, M.~L., {Lequeux}, J., {Maurice}, E., {Prevot}, L., \&
  {Rocca-Volmerange}, B. 1984, \aap, 132, 389

\bibitem[{{Richards} {et~al.}(2006){Richards}, {Lacy}, {Storrie-Lombardi},
  {Hall}, {Gallagher}, {Hines}, {Fan}, {Papovich}, {Vanden Berk}, {Trammell},
  {Schneider}, {Vestergaard}, {York}, {Jester}, {Anderson}, {Budav{\'a}ri}, \&
  {Szalay}}]{Richards06a}
{Richards}, G.~T., {Lacy}, M., {Storrie-Lombardi}, L.~J., {et~al.} 2006, The
  Astrophysical Journal Supplement Series, 166, 470, \dodoi{10.1086/506525}

\bibitem[{{Rieke} {et~al.}(2005){Rieke}, {Kelly}, \& {Horner}}]{Rieke05}
{Rieke}, M.~J., {Kelly}, D., \& {Horner}, S. 2005, in Society of Photo-Optical
  Instrumentation Engineers (SPIE) Conference Series, Vol. 5904, Cryogenic
  Optical Systems and Instruments XI, ed. J.~B. {Heaney} \& L.~G. {Burriesci},
  1--8, \dodoi{10.1117/12.615554}

\bibitem[{{Schlafly} \& {Finkbeiner}(2011)}]{Schlafly11}
{Schlafly}, E.~F., \& {Finkbeiner}, D.~P. 2011, \apj, 737, 103,
  \dodoi{10.1088/0004-637X/737/2/103}

\bibitem[{{Shen} {et~al.}(2011){Shen}, {Richards}, {Strauss}, {Hall},
  {Schneider}, {Snedden}, {Bizyaev}, {Brewington}, {Malanushenko},
  {Malanushenko}, {Oravetz}, {Pan}, \& {Simmons}}]{Shen11}
{Shen}, Y., {Richards}, G.~T., {Strauss}, M.~A., {et~al.} 2011, The
  Astrophysical Journal Supplement Series, 194, 45,
  \dodoi{10.1088/0067-0049/194/2/45}

\bibitem[{{Shen} {et~al.}(2019){Shen}, {Wu}, {Jiang}, {Ba{\~n}ados}, {Fan},
  {Ho}, {Riechers}, {Strauss}, {Venemans}, {Vestergaard}, {Walter}, {Wang},
  {Willott}, {Wu}, \& {Yang}}]{Shen19}
{Shen}, Y., {Wu}, J., {Jiang}, L., {et~al.} 2019, \apj, 873, 35,
  \dodoi{10.3847/1538-4357/ab03d9}

\bibitem[{{Shi} {et~al.}(2022){Shi}, {Kremer}, {Grudi{\'c}},
  {Gerling-Dunsmore}, \& {Hopkins}}]{Shi_2022}
{Shi}, Y., {Kremer}, K., {Grudi{\'c}}, M.~Y., {Gerling-Dunsmore}, H.~J., \&
  {Hopkins}, P.~F. 2022, arXiv e-prints, arXiv:2208.05025.
\newblock \doarXiv{2208.05025}

\bibitem[{{Shibuya} {et~al.}(2015){Shibuya}, {Ouchi}, \&
  {Harikane}}]{Shibuya15}
{Shibuya}, T., {Ouchi}, M., \& {Harikane}, Y. 2015, \apjs, 219, 15,
  \dodoi{10.1088/0067-0049/219/2/15}

\bibitem[{{Stefanon} {et~al.}(2022){Stefanon}, {Bouwens}, {Illingworth},
  {Labb{\'e}}, {Oesch}, \& {Gonzalez}}]{Stefanon_2022}
{Stefanon}, M., {Bouwens}, R.~J., {Illingworth}, G.~D., {et~al.} 2022, \apj,
  935, 94, \dodoi{10.3847/1538-4357/ac7e44}

\bibitem[{{Stefanon} {et~al.}(2017){Stefanon}, {Yan}, {Mobasher}, {Barro},
  {Donley}, {Fontana}, {Hemmati}, {Koekemoer}, {Lee}, {Lee}, {Nayyeri}, {Peth},
  {Pforr}, {Salvato}, {Wiklind}, {Wuyts}, {Ashby}, {Castellano}, {Conselice},
  {Cooper}, {Cooray}, {Dolch}, {Ferguson}, {Galametz}, {Giavalisco}, {Guo},
  {Willner}, {Dickinson}, {Faber}, {Fazio}, {Gardner}, {Gawiser}, {Grazian},
  {Grogin}, {Kocevski}, {Koo}, {Lee}, {Lucas}, {McGrath}, {Nandra}, {Newman},
  \& {van der Wel}}]{Stefanon17}
{Stefanon}, M., {Yan}, H., {Mobasher}, B., {et~al.} 2017, \apjs, 229, 32,
  \dodoi{10.3847/1538-4365/aa66cb}

\bibitem[{{Stevans} {et~al.}(2018){Stevans}, {Finkelstein}, {Wold},
  {Kawinwanichakij}, {Papovich}, {Sherman}, {Ciardullo}, {Florez}, {Gronwall},
  {Jogee}, {Somerville}, \& {Yung}}]{Stevans18}
{Stevans}, M.~L., {Finkelstein}, S.~L., {Wold}, I., {et~al.} 2018, \apj, 863,
  63, \dodoi{10.3847/1538-4357/aacbd7}

\bibitem[{{Tamura} {et~al.}(2019){Tamura}, {Mawatari}, {Hashimoto}, {Inoue},
  {Zackrisson}, {Christensen}, {Binggeli}, {Matsuda}, {Matsuo}, {Takeuchi},
  {Asano}, {Sunaga}, {Shimizu}, {Okamoto}, {Yoshida}, {Lee}, {Shibuya},
  {Taniguchi}, {Umehata}, {Hatsukade}, {Kohno}, \& {Ota}}]{Tamura19}
{Tamura}, Y., {Mawatari}, K., {Hashimoto}, T., {et~al.} 2019, \apj, 874, 27,
  \dodoi{10.3847/1538-4357/ab0374}

\bibitem[{{Tang} {et~al.}(2021){Tang}, {Stark}, {Chevallard}, {Charlot},
  {Endsley}, \& {Congiu}}]{Tang_2021}
{Tang}, M., {Stark}, D.~P., {Chevallard}, J., {et~al.} 2021, \mnras, 503, 4105,
  \dodoi{10.1093/mnras/stab705}

\bibitem[{{Toyouchi} {et~al.}(2022){Toyouchi}, {Inayoshi}, {Li}, {Haiman}, \&
  {Kuiper}}]{Toyouchi_2022}
{Toyouchi}, D., {Inayoshi}, K., {Li}, W., {Haiman}, Z., \& {Kuiper}, R. 2022,
  \mnras, \dodoi{10.1093/mnras/stac3191}

\bibitem[{{Trakhtenbrot} {et~al.}(2011){Trakhtenbrot}, {Netzer}, {Lira}, \&
  {Shemmer}}]{Trakhtenbrot11}
{Trakhtenbrot}, B., {Netzer}, H., {Lira}, P., \& {Shemmer}, O. 2011, \apj, 730,
  7, \dodoi{10.1088/0004-637X/730/1/7}

\bibitem[{{Valiante} {et~al.}(2018){Valiante}, {Schneider}, {Zappacosta},
  {Graziani}, {Pezzulli}, \& {Volonteri}}]{Valiante_2018}
{Valiante}, R., {Schneider}, R., {Zappacosta}, L., {et~al.} 2018, \mnras, 476,
  407, \dodoi{10.1093/mnras/sty213}

\bibitem[{{Vanden Berk} {et~al.}(2001){Vanden Berk}, {Richards}, {Bauer},
  {Strauss}, {Schneider}, {Heckman}, {York}, {Hall}, {Fan}, {Knapp},
  {Anderson}, {Annis}, {Bahcall}, {Bernardi}, {Briggs}, {Brinkmann}, {Brunner},
  {Burles}, {Carey}, {Castander}, {Connolly}, {Crocker}, {Csabai}, {Doi},
  {Finkbeiner}, {Friedman}, {Frieman}, {Fukugita}, {Gunn}, {Hennessy},
  {Ivezi{\'c}}, {Kent}, {Kunszt}, {Lamb}, {Leger}, {Long}, {Loveday}, {Lupton},
  {Meiksin}, {Merelli}, {Munn}, {Newberg}, {Newcomb}, {Nichol}, {Owen}, {Pier},
  {Pope}, {Rockosi}, {Schlegel}, {Siegmund}, {Smee}, {Snir}, {Stoughton},
  {Stubbs}, {SubbaRao}, {Szalay}, {Szokoly}, {Tremonti}, {Uomoto}, {Waddell},
  {Yanny}, \& {Zheng}}]{VB01}
{Vanden Berk}, D.~E., {Richards}, G.~T., {Bauer}, A., {et~al.} 2001, \aj, 122,
  549, \dodoi{10.1086/321167}

\bibitem[{{Walter} {et~al.}(2022){Walter}, {Neeleman}, {Decarli}, {Venemans},
  {Meyer}, {Weiss}, {Ba{\~n}ados}, {Bosman}, {Carilli}, {Fan}, {Riechers},
  {Rix}, \& {Thompson}}]{Walter22}
{Walter}, F., {Neeleman}, M., {Decarli}, R., {et~al.} 2022, \apj, 927, 21,
  \dodoi{10.3847/1538-4357/ac49e8}

\bibitem[{{Wang} {et~al.}(2021){Wang}, {Yang}, {Fan}, {Hennawi}, {Barth},
  {Banados}, {Bian}, {Boutsia}, {Connor}, {Davies}, {Decarli}, {Eilers},
  {Farina}, {Green}, {Jiang}, {Li}, {Mazzucchelli}, {Nanni}, {Schindler},
  {Venemans}, {Walter}, {Wu}, \& {Yue}}]{Wang21}
{Wang}, F., {Yang}, J., {Fan}, X., {et~al.} 2021, \apjl, 907, L1,
  \dodoi{10.3847/2041-8213/abd8c6}

\bibitem[{{Wilkins} {et~al.}(2020){Wilkins}, {Lovell}, {Fairhurst}, {Feng},
  {Matteo}, {Croft}, {Kuusisto}, {Vijayan}, \& {Thomas}}]{Wilkins_2020}
{Wilkins}, S.~M., {Lovell}, C.~C., {Fairhurst}, C., {et~al.} 2020, \mnras, 493,
  6079, \dodoi{10.1093/mnras/staa649}

\bibitem[{{Willott} {et~al.}(2010){Willott}, {Albert}, {Arzoumanian},
  {Bergeron}, {Crampton}, {Delorme}, {Hutchings}, {Omont}, {Reyl{\'e}}, \&
  {Schade}}]{Willott10a}
{Willott}, C.~J., {Albert}, L., {Arzoumanian}, D., {et~al.} 2010, \aj, 140,
  546, \dodoi{10.1088/0004-6256/140/2/546}

\bibitem[{{Wu} {et~al.}(2015){Wu}, {Wang}, {Fan}, {Yi}, {Zuo}, {Bian}, {Jiang},
  {McGreer}, {Wang}, {Yang}, {Yang}, {Thompson}, \& {Beletsky}}]{Wu15}
{Wu}, X.-B., {Wang}, F., {Fan}, X., {et~al.} 2015, \nat, 518, 512,
  \dodoi{10.1038/nature14241}

\bibitem[{{Yang} {et~al.}(2021){Yang}, {Wang}, {Fan}, {Barth}, {Hennawi},
  {Nanni}, {Bian}, {Davies}, {Farina}, {Schindler}, {Ba{\~n}ados}, {Decarli},
  {Eilers}, {Green}, {Guo}, {Jiang}, {Li}, {Venemans}, {Walter}, {Wu}, \&
  {Yue}}]{Yang21}
{Yang}, J., {Wang}, F., {Fan}, X., {et~al.} 2021, \apj, 923, 262,
  \dodoi{10.3847/1538-4357/ac2b32}

\bibitem[{{Yuan} \& {Narayan}(2014)}]{Yuan_Narayan_2014}
{Yuan}, F., \& {Narayan}, R. 2014, \araa, 52, 529,
  \dodoi{10.1146/annurev-astro-082812-141003}

\bibitem[{{Zakamska} {et~al.}(2003){Zakamska}, {Strauss}, {Krolik}, {Collinge},
  {Hall}, {Hao}, {Heckman}, {Ivezi{\'c}}, {Richards}, {Schlegel}, {Schneider},
  {Strateva}, {Vanden Berk}, {Anderson}, \& {Brinkmann}}]{Zakamska_2003}
{Zakamska}, N.~L., {Strauss}, M.~A., {Krolik}, J.~H., {et~al.} 2003, \aj, 126,
  2125, \dodoi{10.1086/378610}

\end{thebibliography}
\bibliographystyle{aasjournal}

\end{document}